\documentclass[symmetry,article,accept,moreauthors]{Definitions/mdpi}

\firstpage{1} 
\pubvolume{1}
\issuenum{1}
\articlenumber{0}
\pubyear{2026}
\copyrightyear{2026}
\externaleditor{Vasilis K. Oikonomou} 
\datereceived{6 October 2026} 
\daterevised{26 October 2026} 
\dateaccepted{28 October 2026} 
\datepublished{ } 

\graphicspath{{Definitions/}}
\usepackage{longtable} 
\usepackage{comment}
\usepackage{rotating}
\usepackage{float}

\newcommand{\aj}{Astron. J.}
\newcommand{\apj}{Astrophys. J.}

\newcommand{\apjs}{Astrophys. J. Suppl. Ser.}

\newcommand{\araa}{Annu. Rev. Astron. Astrophys.}

\newcommand{\aap}{Astron. Astrophys.}
\newcommand{\aaps}{Astron. Astrophys. Suppl.}

\newcommand{\pasp}{Publ. Astron. Soc. Pac.}
\newcommand{\pasj}{Publ. Astron. Soc. Jpn.}
\newcommand{\mnras}{Mon. Not. R. Astron. Soc.}

\newcommand{\aapr}{Astron. Astrophys. Rev.}
\newcommand{\baas}{Bull. Am. Astron. Soc.}

\Title{The Peculiar Astrometric Behaviour and Radio Structure of the~Quasar 1524$-$136  Revealed with Very Long Baseline~Interferometry}

\Author{Angelina Osetrova $^{1}$\orcidA{}, S\'andor Frey $^{2,3,}$*\orcidB{}, Oleg Titov $^{4,5 }$\orcidC{}, Alexey Melnikov $^{1}$\orcidD{} and Panna Gelle $^{2,6}$\orcidE{}}

\AuthorNames{Firstname Lastname, Firstname Lastname and Firstname Lastname}

\address{%
$^{1}$ \quad Institute of Applied Astronomy, Russian Academy of Sciences, Kutuzova  Embankment 10, 191187~St.~Petersburg, Russia; aa.osetrova@iaaras.ru (A.O.); aem@iaaras.ru (A.M.)\\
$^{2}$ \quad Konkoly Observatory, HUN-REN Research Centre for Astronomy and Earth Sciences (CSFK), MTA Centre of Excellence, Konkoly Thege Mikl\'{o}s \'{u}t 15-17, 1121~Budapest, Hungary; gellepanna@student.elte.hu\\
$^{3}$ \quad Department of Astronomy, Institute of Physics and Astronomy, ELTE E\"otv\"os Lor\'and University, P\'azm\'any P\'eter s\'et\'any 1/A, 1117~Budapest, Hungary\\
$^{4}$ \quad Shanghai Astronomical Observatory, Chinese Academy of Sciences, 80 Nandan Road, Shanghai~200030, China; oleg.titov@shao.ac.cn\\ 
$^{5}$ \quad Phase\&Rate, Canberra~2600, Australia\\
$^{6}$ \quad Institute of Physics and Astronomy, ELTE E\"otv\"os Lor\'and University, P\'azm\'any P\'eter s\'et\'any 1/A, 1117~Budapest, Hungary
}

\corres{Correspondence: frey.sandor@csfk.org}

\abstract{According to astrometric very long baseline interferometry (VLBI) observations accurate to sub-milliarcsecond levels, the radio quasar 1524$-$136 (J1526$-$1351) exhibited an exceptionally large shift, more than $0.1^{\prime\prime}$, between the 1990s and 2010s. This luminous quasar is known for its extended ($\sim0.4^{\prime\prime}$), highly asymmetric and non-collinear jet structure. Here we analyse historical VLBI astrometric and imaging data, as well as \textit{Gaia} optical astrometric information, to find the cause of the sudden {apparent} change in the position. At the $X$ band ($\sim8.4$~GHz frequency), where radio positions are conventionally measured, the core region brightened substantially relative to the hotspot at the end of the northern jet. As a result, the brightness peak shifted from the hotspot to the core, causing the VLBI position to shift accordingly. Similar changes could affect the astrometric positions of other radio sources. A better understanding of the astrophysical reasons behind them is important for maintaining the stability of the VLBI-based celestial reference frame.}


\begin{document}

\section{Introduction}
\label{sect:intro}

Radio-loud (jetted) active galactic nuclei (AGN) are powered by accretion onto supermassive black holes located in the centres of galaxies. Relativistic jets that produce radio emission via the synchrotron mechanism are among the most extreme astrophysical environments in the Universe~\cite{2019ARA&A..57..467B}. They transport energy and momentum from the vicinity of the central black hole to the surrounding interstellar and even intergalactic medium. This feedback process influences star formation, gas cooling, and~the evolution of galaxies, groups, and~clusters~\cite{2012ARA&A..50..455F,2025Galax..13..102M}. Radio emission traces relativistic plasma on spatial scales through several orders of magnitude, from~sub-pc to Mpc. Radio-loud AGN serve as natural laboratories to study jet launching and collimation, accretion physics, particle acceleration, magnetic fields, and~the interaction of jets with their environment. Owing to their enormous power, jetted AGN can be detected from vast distances, potentially making them useful probes of cosmological models~\cite{2026A&A...710A.133G}. 

Apart from their high relevance for astrophysics and cosmology, jetted AGN are also important objects for astrometry. Distant compact extragalactic radio sources can provide stable reference points on the sky, forming a quasi-inertial reference system. Using the technique of very long baseline interferometry (VLBI), not only compact jets can be imaged with unmatched resolution~\cite{2019Galax...8....1H}, but~it is also possible to locate jet emission features with sub-milliarcsecond (sub-mas) precision~\cite{1998RvMP...70.1393S}. This provides the basis of the International Celestial Reference Frame (ICRF) \cite{2022Univ....8..374D}. However, AGN in general are not ideally point-like and positionally stable sources. Frequency-dependent core shifts~\cite{2011A&A...532A..38S}, evolving jets~\cite{2011AJ....141..178M}, and structural changes in their brightness distribution~\cite{2022MNRAS.512..874T, Osetrova2024} can produce apparent changes in their position. Understanding these effects is essential for improving the accuracy and stability of the VLBI reference~frame.

Modern optical astrometry with the \textit{Gaia} spacecraft~\cite{2016A&A...595A...1G} provides AGN positions with accuracy comparable to that of the VLBI-based ICRF. Comparing optical and radio positions~\cite{2013A&A...553A..13O} and quantifying apparent offsets is a powerful tool for improving the accuracy of the radio--optical link of reference frames~\cite{2016A&A...595A...5M,2017MNRAS.467L..71P} and for studying AGN structure and the physical connection between the radio- and optical-emitting regions~\cite{2017A&A...598L...1K,2019ApJ...871..143P,2024A&A...684A.202L,2025MNRAS.543..479P}.

Here we present a case study of a luminous radio quasar, 1524$-$136 ({PKS~1524$-$13, OR~$-$140,} J1526$-$1351), which was identified as an AGN with significant apparent radio--optical positional offset exceeding $\sim100$~mas before the \textit{Gaia} era~\cite{2013A&A...553A..13O}. By~analysing and interpreting VLBI astrometric and imaging data published in the literature and available in the archives, we trace back the anomalous astrometric behaviour of 1524$-$136 to a substantial change in its brightness distribution that must have happened some time between 1998 and 2014. In~Section~\ref{sect:target}, we introduce the target of our study. Section~\ref{sect:astrometry} reports on the astrometric measurements available. Section~\ref{sect:imaging} presents the results of selected VLBI imaging observations. The~interpretation of the astrometric results in the context of the changes in the radio structure is discussed in Section~\ref{sect:disc}. Finally, our conclusions are given in Section~\ref{sect:concl}.

\section{The Target~Object}
\label{sect:target}

The quasar 1524$-$136 at redshift $z=1.687$ \cite{1993ApJS...87..451H} was identified as a strong radio source with $1.4$ GHz flux density $S_{1.4}=2.7$~Jy and a steep power-law spectrum with $\alpha=-0.6$ spectral index (defined as $S \propto \nu^\alpha$, where $\nu$ is  observing frequency) {measured in the frequency range $0.4-2.6$~GHz} already in the 1960s~\cite{1966AuJPh..19..559B}. It was subsequently classified as a compact steep-spectrum (CSS) radio quasar~\cite{1992MNRAS.257..353M,1995A&AS..113..409S}. CSS quasars are powerful AGN with a subgalactic extent ($\lesssim20$~kpc) and steep spectrum in the GHz frequency regime ($\alpha<-0.5$). \textls[-15]{CSS sources are usually interpreted as young ($\sim10^{4-5}$~yr old) jetted AGN whose relativistic plasma jets emanating from the vicinity of the central supermassive black hole in opposite directions did not yet propagate outside their host galaxy. However, in~an alternative scenario called frustration, the~jets are not necessarily very young, but their expansion is obstructed by the dense interstellar medium which confines them to within the host galaxy~\cite{2021A&ARv..29....3O}. }

The ``compactness'' of CSS sources is to be understood at the arcsecond angular scales probed by connected-element interferometers like the Karl G. Jansky Very Large Array (VLA). Indeed, 1524$-$136 is only slightly extended at $5$ and $8.4$~GHz and~resolved into two components separated  by $\lesssim400$~mas at $15$~GHz with the VLA~\cite{1994A&A...292...59M,1999ApJS..124..285R}. In the standard flat $\Lambda$ Cold Dark Matter cosmological model with $\Omega_{\mathrm m} = 0.3$, $\Omega_{\Lambda} = 0.7$, and~$H_0 = 70$~km\,s$^{-1}$\,Mpc$^{-1}$, $1^{\prime\prime}$ angular size corresponds to $8.466$~kpc projected linear size at $z = 1.687$ \cite{2006PASP..118.1711W}.

\begin{figure}[H]

\includegraphics[width=9.2 cm]{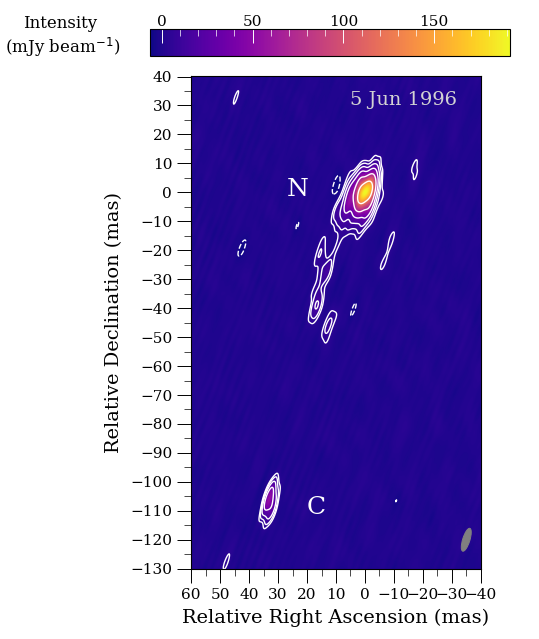}
\caption{The $4.9$ GHz VLBA image of 1524$-$136 from the VLBI Space Observatory Programme prelaunch survey~\cite{2000ApJS..131...95F} conducted in 1996, reprocessed here with a wider field of view to reveal the extended radio structure. The~northern hotspot and the core are labelled as N and C, respectively, as~identified based on other VLBI observations in the 1990s~\cite{2002A&A...389L..15M,2002A&A...389...58M}. The~peak intensity is $192$~mJy\,beam$^{-1}$. The~lowest contours are drawn at $\pm4$~mJy\,beam$^{-1}$. The~positive contour levels increase by a factor of $2$. The~full width at half-maximum (FWHM) of the elliptical Gaussian restoring beam is $8.27\,\mathrm{mas} \times 2.5\,\mathrm{mas}$ at major axis position angle $\mathrm{PA} = -17.4^{\circ}$, as~indicated in the bottom right corner. The~coordinates are relative to the phase centre whose position was determined with $\sim10$~mas accuracy as right ascension $15^{\mathrm{h}}\,26^{\mathrm{min}}\,59.440^{\mathrm{s}}$ and declination $-13^{\circ}\,51^{\prime}\,00.17^{\prime\prime}$ \cite{2000ApJS..131...95F} (Table~\ref{tab:pos}). \label{fig:c-1996}} 
\end{figure}

When imaged with VLBI at mas-scale resolution, CSS quasars show a variety of morphologies, typically displaying multiple components, asymmetric two-sided, one-sided core--jet, or~complex structures (e.g., \cite{2007A&A...469..437K,2013MNRAS.433..147D}). The~first published VLBI image of 1524$-$136 was obtained in 1996 as part of a $4.9$ GHz snapshot survey of bright radio AGN with the Very Long Baseline Array (VLBA) \cite{2000ApJS..131...95F}. While that image showed a single but resolved feature, {our} re-analysis of the same data by opening a larger field of view uncovered a nearly linear multi-component structure extending to $\sim100$~mas (Figure~\ref{fig:c-1996}). Here the brightest, northernmost component is a terminal hotspot at the end of the northern jet. More sensitive 1.7 GHz global VLBI, 8.4 GHz VLBA~\cite{2002A&A...389L..15M}, and~$5$ GHz VLBA imaging~\cite{2002A&A...389...58M} in 1992, 1996, and~1998, respectively, revealed a quite unique structure also featuring  a faint and extended ($\sim280$~mas) counter-jet pointing towards the southwest. {However, these were not astrometric or phase-referenced imaging observations; therefore  information on the absolute astrometric position was not preserved.} The two-sided radio jet structure is highly non-collinear and asymmetric. The~northern jet {at the approximate position angle $-15^{\circ}$} appears significantly brighter and ends closer to the nucleus than the southwestern counter-jet {at the position angle $-135^{\circ}$. (Position angles are conventionally measured from north to east.) The brightness asymmetry and misalignment of the jet and the counter-jet suggest that frustration is at play in this CSS source.} The core was identified with a barely resolved, weakly polarised~\cite{2002A&A...389...58M} component in a central location. Notably, the~brightness peak fell on the northern hotspot and not the core at all frequency bands in the 1990s (Figure~\ref{fig:c-1996}) \cite{2000ApJS..131...95F,2002A&A...389L..15M,2002A&A...389...58M}.

{In the optical position, the~\textit{Gaia} \cite{2016A&A...595A...1G} Data Release 3 (DR3,~\cite{2023AA...674A...1G}) lists 1524$-$136 with $G$-band mean magnitude 19.7. The~formal positional uncertainties both in right ascension ($\alpha$) and declination ($\delta$) are $0.41$~mas, with~an astrometric excess noise of $0.7$~mas, albeit with low statistical significance (0.4), indicating an astrometrically stable object.}

\begin{table}[H]
\caption{VLBI and optical position determinations for 1524$-$136 in the~literature.\label{tab:pos}}
\small
\begin{tabularx}{\textwidth}{lllclllcCC}
\toprule
\multicolumn{3}{c}{\boldmath{$\alpha$}}	& \boldmath{$\Delta\alpha$}	& \multicolumn{3}{c}{\boldmath{$\delta$}}     & \boldmath{$\Delta\delta$} & \textbf{Epoch} & \textbf{Ref.} \\
\textbf{[h]} & \textbf{[min]} & \textbf{[s]}   & \textbf{[mas]} & \boldmath{[$^{\circ}$]} & \boldmath{[$^{\prime}$]} & \boldmath{[$^{\prime\prime}$]}  & \textbf{[mas]} & \textbf{[yr]} &   \\
\multicolumn{3}{c}{\textbf{(1)}} & \textbf{(2)} & \multicolumn{3}{c}{\textbf{(3)}} & \textbf{(4)} & \textbf{(5)} & \textbf{(6)} \\            
\midrule
$15$ & $26$ & $59.440$   & $10$  & $-13$ & $51$ & $00.17$   & $10$ & $1996.427$ & \citep{2000ApJS..131...95F} * \\
$15$ & $26$ & $59.44070$ & $2.4$ & $-13$ & $51$ & $00.1638$ & $3.9$ & $2000.0$   & ICRF2~\cite{2015AJ....150...58F} {*}\\
$15$ & $26$ & $59.44274$ & $0.2$ & $-13$ & $51$ & $00.2644$ & $0.5$ & $2015.0$   & ICRF3~\cite{2020AA...644A.159C}\\
\midrule
$15$ & $26$ & $59.44260$ & $0.4$ & $-13$ & $51$ & $00.2572$ & $0.4$ & $2016.0$   & \textit{Gaia} DR3~\cite{2023AA...674A...1G} \\

\bottomrule
\end{tabularx}
\noindent{\footnotesize{Notes: {* These positions refer to the northern hotspot, while the others refer to the core (see the text for more details).} Col.~1---right ascension; Col.~2---right ascension formal uncertainty; Col.~3---declination; Col.~4---declination formal uncertainty; Col.~5---coordinate {reference} epoch; Col.~6---literature reference}.}
\end{table}

\section{Astrometric Observations and~Analysis}
\label{sect:astrometry}
 
The radio source 1524$-$136 has been observed in geodetic/astrometric VLBI programmes coordinated by the International VLBI Service for Geodesy and Astrometry (IVS) \cite{2017JGeod..91..711N}, primarily with the VLBA at the $S/X$ bands ($2.3/8.4$~GHz), since 1997, although~these observations were not regular. The~position of the source in the second realisation of the ICRF (ICRF2,~\cite{2015AJ....150...58F}) was determined from a single VLBI observing session conducted in 1997 (Table~\ref{tab:pos}). Beginning in 2014, the~source was observed in six additional $S/X$-band VLBA~sessions.

The position derived from the 1997 observation shows a significant offset relative to the cluster of positions obtained after 2014. During~the construction of the next and current ICRF realisations (ICRF3,~\cite{2020AA...644A.159C}), this early measurement was considered likely to be affected by an error. Consequently, the~ICRF3 position of 1524$-$136 (Table~\ref{tab:pos}) was determined exclusively from three sessions of the post-2014~observations. 
 
In addition, 1524$-$136 was included in the Australia Telescope National Facility (ATNF) observing programme (project code: V515), aimed at improving the celestial reference frame by geodetic VLBI observations of Southern Hemisphere radio sources. For~this project, nine~radio sources with large astrometric position displacements (>5~mas) were selected for geodetic VLBI observations using single-frequency mode in the $X$ band.

The $24$ h experiment V515E was carried out on 11--12 February 2022 using a radio telescope network consisting of $11$ stations: a $22$ m antenna of the Australia Telescope Compact Array (ATCA), Parkes ($64$~m), Morpa ($22$~m), Ceduna ($30$~m), Hobart ($12$~m), Tidbinbilla (70~m), Yarragadee (AuScope $12$ m antenna) in Australia; Warkworth $30$ m antenna operated by SpaceOps NZ in New Zealand; Hartebeesthoek $26$ m antenna of the South African Radio Astronomy Observatory; and $32$ m Svetloe and Zelenchukskaya {antennas} of the Quasar VLBI Network operated by the Institute of Applied Astronomy in~Russia. 

The observational schedule was prepared in a standard geodetic manner where observations of the target sources are mixed with the calibrator scans in varying durations from $120$ to $1300$~s of observing time, depending on the a priori flux density of the radio sources. The~phase and the fringe calibrator sources were observed  every 1--2~h and 3--4~h, respectively; however, all of them were included in a further astrometric~analysis.

A single $X$-band frequency setup was used for several purposes, to increase sensitivity to the faint target sources and~to accommodate some of the {Southern Hemisphere} Long Baseline Array (LBA) facilities equipped with a single-band receiver, which have therefore have never been used before for geodetic/astrometric VLBI observations, including the 30 m radio telescopes in Warkworth and Ceduna. The~frequency setup for all {antennas} consists of a number of intermediate frequency channels (IFs) and of various bandwidths and polarisations. Information on the basic setup of the {antennas} is presented in  Table~\ref{tab:v515e}. 

\begin{table}[H]
\caption{V515E experiment {(11--12 February 2022)} frequency setup at  participating~stations.}
\label{tab:v515e}
\begin{tabularx}{\textwidth}{LCCC}
\toprule
\textbf{Station} & \textbf{Frequency} & \textbf{Polarisation} &
\textbf{IF Number} \boldmath{$\times$} \\
\textbf{Name}    & \textbf{Range}     &                       & 
\textbf{IF Bandwidth}       \\
                 & \textbf{[MHz]}     &                       & 
\textbf{[MHz]}              \\
\midrule
ATCA           & 8198--8598 & RCP *  & $2 \times 64$ \\
Parkes         & 8198--8598 & RCP  & $2 \times 64$ \\
Morpa          & 8198--8598 & RCP  & $2 \times 64$ \\
Tidbinbilla    & 8198--8598 & RCP  & $8 \times 16$ \\
Yarragadee     & 8198--8930 & RCP  & $16 \times 16$ \\
Ceduna         & 8198--8930 & RCP  & $16 \times 16$ \\
Warkworth      & 8198--8930 & RCP  & $16 \times 16$ \\
Hartebeesthoek & 8198--8930 & RCP  & $16 \times 16$ \\
Svetloe        & 8198--8930 & RCP  & $8 \times 32$ \\
Zelenchukskaya & 8198--8930 & RCP  & $8 \times 32$ \\
Hobart         & 8198--8930 & HV **  & $16 \times 32$ \\
\bottomrule
\end{tabularx}

\noindent{\footnotesize{* right circular polarisation; ** dual linear polarisation, {converted to circular polarisation using PolConvert~\cite{2016A&A...587A.143M}}}}
\end{table}

All participating radio telescopes were equipped with sensitive cooled receiving systems and standard VLBI signal registration hardware. After~the observations were finished, the~recorded data were sent to the Pawsey Supercomputing Research Centre via the Internet and correlated with the DiFX software correlator~\cite{2011PASP..123..275D} using typical spectral and time resolution parameters. The~correlated dataset was analysed with the Haystack Observatory Postprocessing System (HOPS) (software available at \url{https://www.haystack.mit.edu/haystack-observatory-postprocessing-system-hops/}, accessed on 5 October 2026).

\subsection*{Positions at $S$ and $X$ Bands}
\label{subsec:sx}

According to the standard geodetic VLBI approach~\cite{1998RvMP...70.1393S}, radio sources are observed simultaneously at two frequency bands, $S$ and $X$. The~reported astrometric position conventionally refers to the $X$ band, while the $S$-band observations are primarily used to calibrate the dispersive ionospheric delay by forming the ionosphere-free linear combination described below. The~coordinates obtained this way are referred to as $S/X$. The~ionospheric delay at  frequency $f$ is 
\vspace{-3pt}
\begin{equation}
    \tau_{\mathrm{ion}, f} \approx 40.3 \frac{E_\mathrm{{s}}}{cf^2}  \approx  \frac{b}{f^2},
    \label{eq:tau_ion}
\end{equation}
where $b = 40.3 \frac{E_\mathrm{{s}}}{c}$ is the parameter reflecting the proportional dependence of the ionospheric correction on  observation frequency, $c$ is the speed of light, and~$E_\mathrm{{s}}$ (slant total electron content, sTEC) is an integral quantity representing the total number of  free electrons along the path of a radio signal propagating through the ionosphere to the radio telescope from the target source (e.g.,~\cite{IERSTN36}). The~value of $E_\mathrm{{s}}$ cannot be estimated from a single-band VLBI observation alone, only from dual-band observations or from external~data.

The delay for the radio source at the $X$ and $S$ bands can be expressed by the following system of equations:
\begin{equation}
\begin{cases}
	\tau_X &=  \tau_{0,X}  + \tau_{\mathrm{ion}, X} = \tau_{0, X}   + \frac{b}{f_X^2} \\
	\tau_S &= \tau_{0,S}   + \tau_{\mathrm{ion}, S} =  \tau_{0, S}   + \frac{b}{f_S^2}.
    \end{cases}
    \label{eq:tau}
\end{equation}
Here $\tau_{0,X}$ and $\tau_{0,S}$ are the ionosphere-free group delays at the $X$ and $S$ bands, respectively. In~this approach, it is implicitly assumed that the source positions at the $S$ and $X$ bands do not differ significantly, so  $ \tau_{0,X}  \approx  \tau_{0,S} \approx  \tau_{0}$.
\begin{equation}
\begin{cases}
	\tau_X &   = \tau_{0}  + \frac{b_{S/X}}{f_X^2} \\
	\tau_S &   =  \tau_{0}  + \frac{b_{S/X}}{f_S^2},
    \end{cases}
    \label{eq:tau0}
\end{equation}
\begin{equation}
    	\tau_X  - \tau_S  = 
        \frac{b_{S/X}}{f_X^2} - \frac{b_{S/X}}{f_S^2}   \Rightarrow 
        b_{S/X} = \frac{\tau_X  - \tau_S}{\frac{1}{f_X^2} - \frac{1}{f_S^2 }}.
      \label{ion:1}
  \end{equation}
Here $b_{S/X} \equiv b$ if the assumption of the identical source positions at the $S$ and $X$ bands holds. 
The ionospheric delay at $X$ band can be expressed in the following form by substituting $b_{S/X}$ from Equation~(\ref{ion:1}) to Equation~(\ref{eq:tau0}), also using Equation~(\ref{eq:tau}), again with the assumption that the radio source coordinates at the $S$ and $X$ bands do not differ from each other:
\begin{equation}
\tau_{\mathrm{ion}, X}  =  (\tau_X  - \tau_S  )\frac{f_S^2}{f_S^2-  f_X^2}.  
 \label{eq:tau_ion_x}
\end{equation}

However, if~a substantial frequency-dependent position offset is present, then $\tau_{0,X}\neq \tau_{0,S}$, and~the transition between  Equations~(\ref{eq:tau}) and  (\ref{eq:tau0}) is invalid. As~a result, $b \neq b_{S/X}$, and~the combined $S/X$ position solution that is supposed to refer to the $X$ band will be~biased.

An alternative method for correcting ionospheric delays employs global ionosphere maps (GIMs) derived from Global Navigation Satellite Systems (GNSS) observations by the Center for Orbit Determination in Europe (CODE) Analysis Center~\cite{Schaer1996}.
In this case, the~ionospheric delay is corrected independently at each VLBI frequency band without assuming the equality of radio source coordinates. The~details of the algorithm are described in~\cite{2019ApJ...885..159A}. The~daily GIMs contain the vertical total content in TEC units (TECU). By~taking into account the positions of the radio source and the radio telescopes, one can estimate sTEC and therefore the time delay correction caused by the ionosphere (Equation~(\ref{eq:tau_ion})).

Using the least-squares method with primary troposphere calibration implemented in the OCCAM (version 6.3) software package~\cite{2004ivsg.conf..267T}, the~{brightness peak} position of 1524$-$136 was estimated in three different~ways:
\begin{itemize}
    \item $S/X$---the conventional combined $X$-band position obtained using the standard dual-frequency ionospheric calibration (Equation~(\ref{eq:tau_ion_x})); 
    \item $S$---the $S$-band position obtained after applying ionospheric corrections derived from GIMs;
    \item $X$---the $X$-band position obtained after applying ionospheric corrections derived from GIMs.
\end{itemize}

The VLBI {right ascension and declination time series plots along with the \textit{Gaia} \cite{2016A&A...595A...1G} DR3~\cite{2023AA...674A...1G}} optical coordinates (Table~\ref{tab:pos}) are presented in Figure~\ref{fig:pos}, relative to $\alpha_0=15^\mathrm{h} 26^\mathrm{min} 59.4427^\mathrm{s}$ and $\delta_0=-13^\circ 51^{\prime} 00.2637^{\prime\prime}$. {This reference position corresponds to the coordinates obtained through a global $S/X$ solution in OCCAM 6.3 based on VLBI observations during the compilation of the source catalogue.} {Note that the different errors of position solutions, especially visible at the $S$ band, are related to the sensitivity of the given observations, determined by the bandwidth and the observing time.} 

\begin{figure}[H]
\includegraphics[width=12cm]{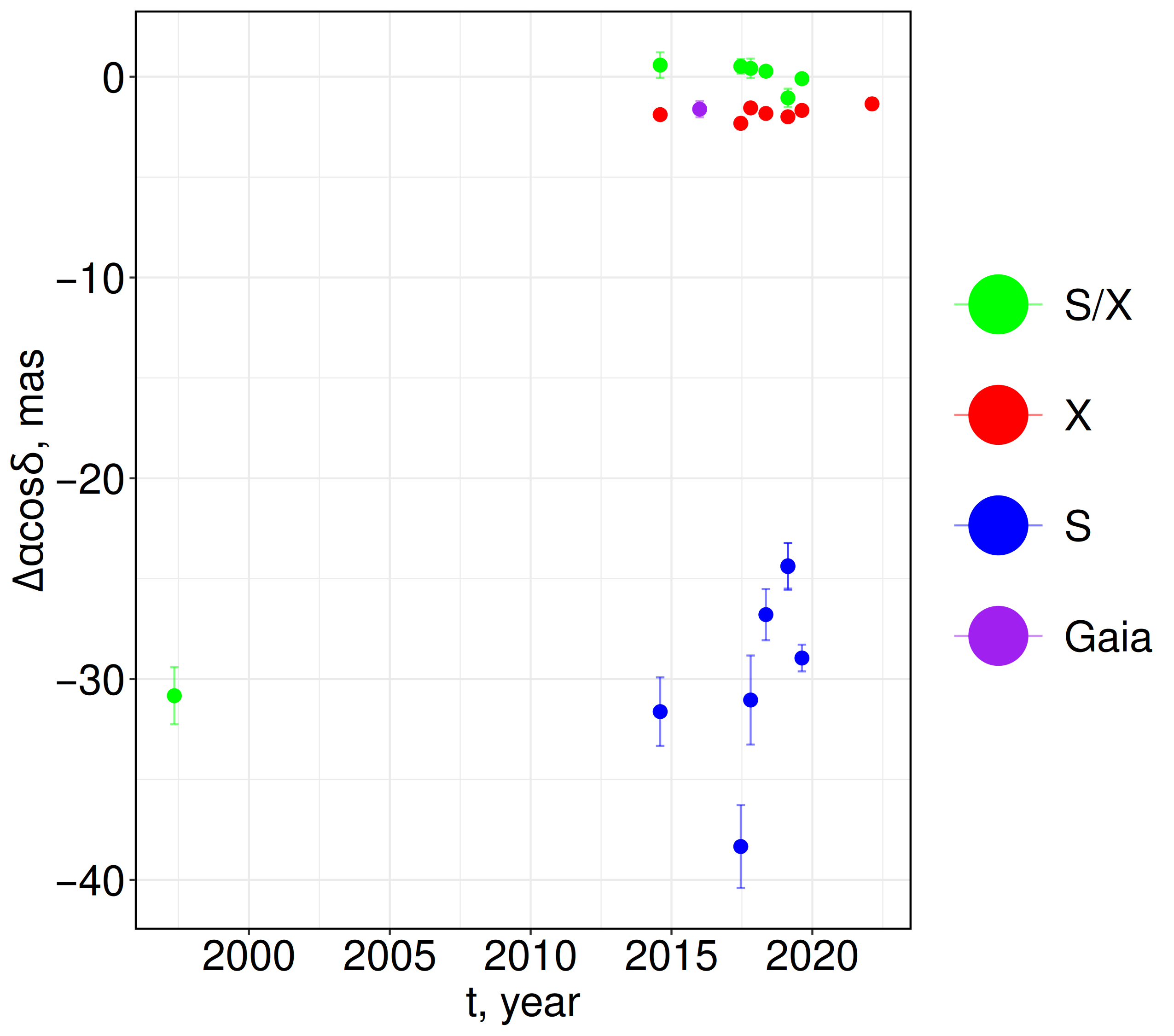} 
\includegraphics[width=12cm]{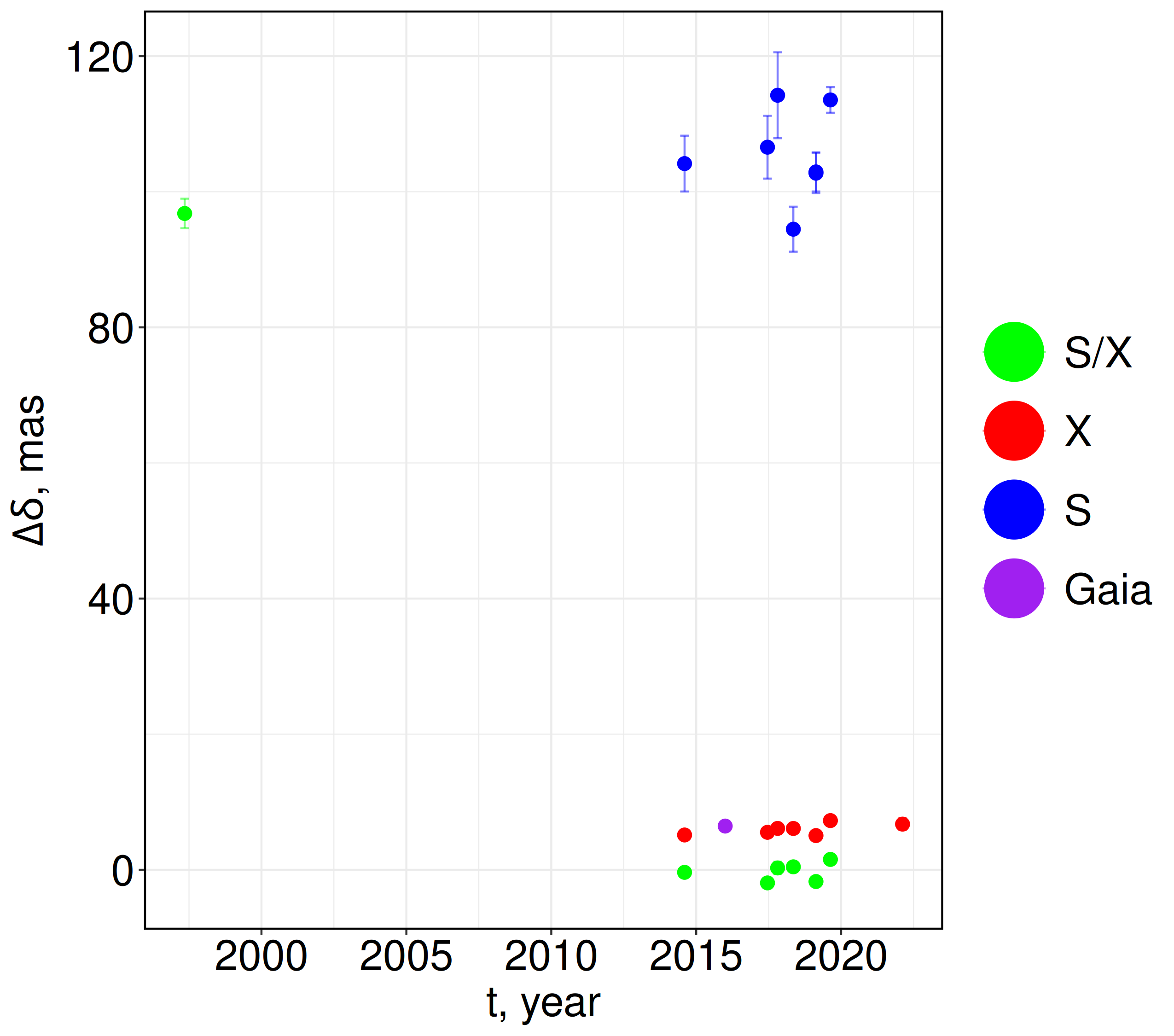}   
\caption{Time series of  corrected right ascension relative to $15^\mathrm{h} 26^\mathrm{min} 59.4427^\mathrm{s}$ (\textbf{top}) and declination relative to $-13^\circ 51^{\prime} 00.2637^{\prime\prime}$ (\textbf{bottom}) VLBI coordinates of 1524$-$136 in three cases: dual-band $S/X$ position (green), $X$-band position with GIM calibration (red), and~$S$-band position with GIM calibration (blue). ~\textit{Gaia} DR3 optical position is also indicated in purple~colour.}\label{fig:pos}
\end{figure}
\unskip   
 
\section{VLBI~Imaging}
\label{sect:imaging}

Archival calibrated VLBI visibility data of 1524$-$136 are publicly available for imaging in the Astrogeo database (\url{https://astrogeo.org/cgi-bin/imdb_get_source.csh?source=J1526-1351}, accessed on 5 October 2026) \cite{2025ApJS..276...38P}. These cover the period between 1996 and 2019, but~with a $\sim17$ yr long gap after 1997, and~include the $C$-band (4.9 GHz) VLBA dataset from 1996~\cite{2000ApJS..131...95F} (Figure~\ref{fig:c-1996}), as~well as the seven epochs of $S/X$-band VLBA data from 1997 and 2014--2018 whose astrometric analysis led to the ICRF2 and ICRF3 positions, respectively (see Section~\ref{sect:astrometry}).

Here we present examples of $S$- and $X$-band VLBI images of 1524$-$136 \mbox{(Figures~\ref{fig:s-1997-2018} and \ref{fig:x-1997-2018}),} to~demonstrate how its mas-scale radio structure changed between the 1990s and the 2010s. These short snapshot observations were not sufficiently sensitive to recover the faint counter-jet known from earlier VLBI observations~\cite{2002A&A...389L..15M} but~characterise the core and the northern jet emission. The~observing information related to the 1997 and 2018--2019 datasets presented here is collected in Table~\ref{tab:obs}. Standard hybrid mapping was performed in  Difmap~\cite{1994BAAS...26..987S} software (version 2.5q). The~image parameters are given in Table~\ref{tab:img}.

\begin{table}[H]
\caption{Archival $S/X$-band VLBA imaging observations of 1524$-$136 analysed~here.\label{tab:obs}}
\begin{adjustwidth}{-\extralength}{0cm}
\begin{tabularx}{\fulllength}{CCCCCC}
\toprule
\textbf{Mean Epoch}	& \textbf{Project Code}	& \textbf{Frequency}  & \textbf{Bandwidth} & \textbf{On-Source Time} & \textbf{Ref.} \\
\textbf{[yr]} &   & \textbf{[GHz]}   & \textbf{[MHz]} & \textbf{[s]} &   \\
\midrule
$1997.344$ & BB023 & $2.27$   & $32$  & $303$ & \citep{2002ApJS..141...13B} \\
 & & $8.34$ & $32$  & $303$ &   \\
$2019.124$ * & BP222 & $2.28$   & $128$  & 20,886 &  \citep{2025ApJS..276...38P} \\
 & & $8.65$   & $384$  & 20,886 &   \\
\bottomrule
\end{tabularx}
\end{adjustwidth}
\noindent{\footnotesize{* A combination of observations made on 11 December 2018, 26 December 2018, and~23 April  2019}.}
\end{table}
\unskip

\begin{figure}[H]

\includegraphics[width=9.8 cm]{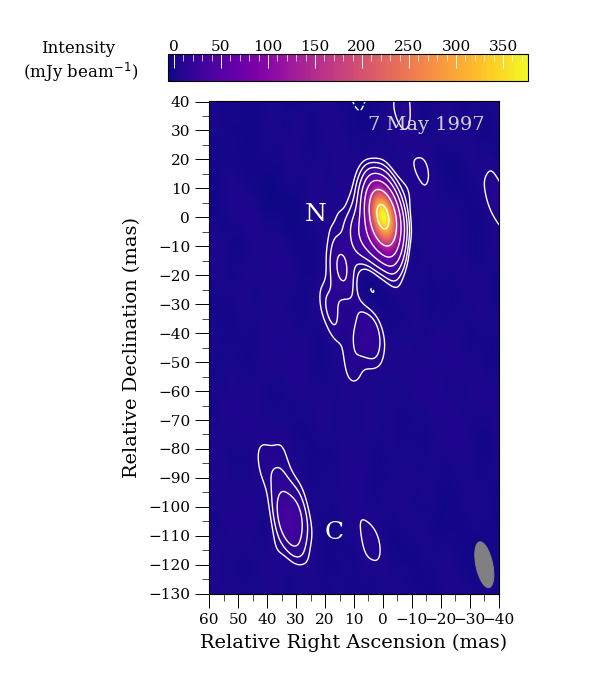}

\caption{\textit{Cont.}}
\end{figure}

\begin{figure}[H]\ContinuedFloat

\includegraphics[width=9.8 cm]{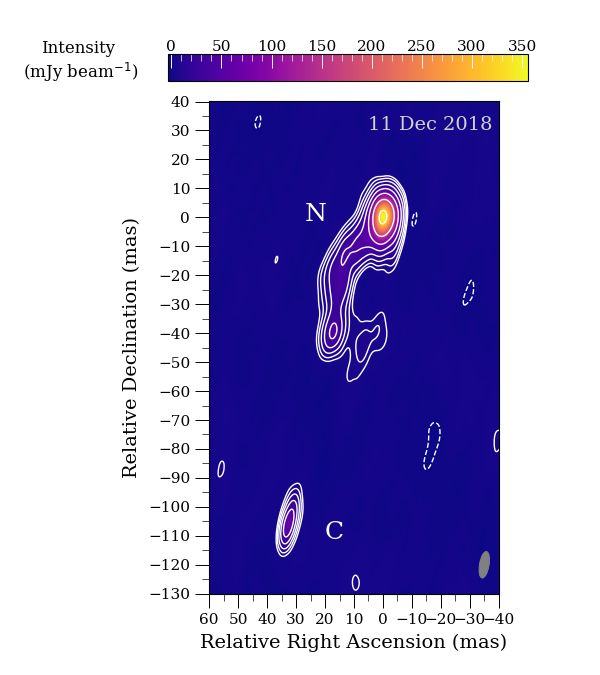}
\caption{Examples of $S$-band VLBI images of 1524$-$136 from 1997 (\textbf{top}) and 2018--2019 (\textbf{bottom}) using archival visibility data obtained from the Astrogeo website. The~field of view and the labelling are the same as in Figure~\ref{fig:c-1996}. The~lowest contours are drawn at about $\pm 3\sigma$ image noise. The~positive contour levels increase by a factor of $2$. The~image parameters are given in Table~\ref{tab:img}. \label{fig:s-1997-2018}} 
\end{figure}
\unskip

\begin{figure}[H]

\includegraphics[width=9.8 cm]{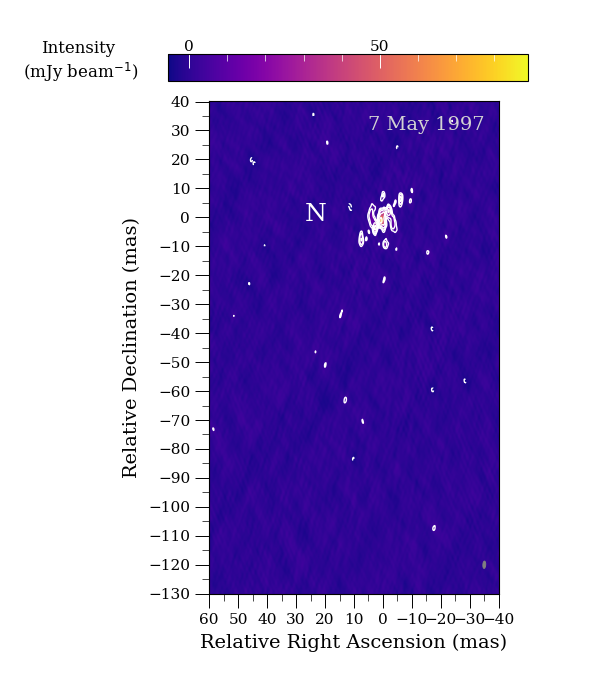}
\caption{\textit{Cont.}}
\end{figure}

\begin{figure}[H]\ContinuedFloat

\includegraphics[width=9.8 cm]{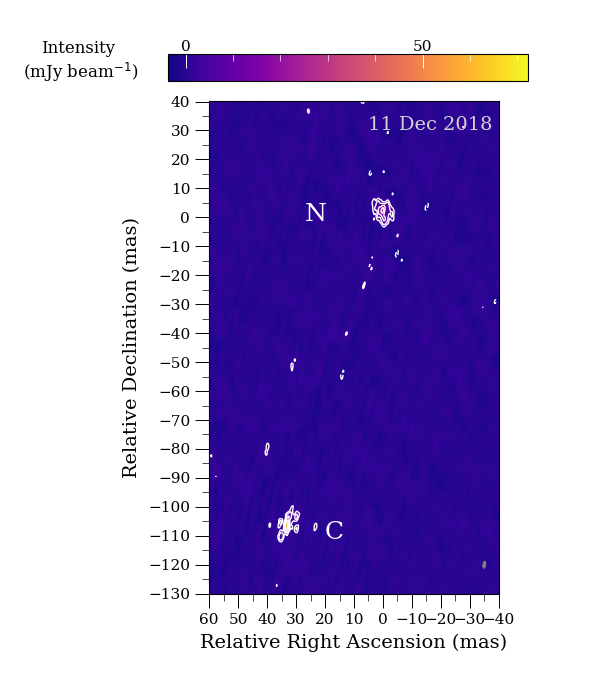}
\caption{Examples of $X$-band VLBI images of 1524$-$136 from 1997 (\textbf{top}) and 2018--2019 (\textbf{bottom}) using archival visibility data obtained from the Astrogeo website. The~field of view and the labelling are the same as in Figure~\ref{fig:c-1996}. The~lowest contours are drawn at about $\pm 3\sigma$ image noise. The~positive contour levels increase by a factor of $2$. The~image parameters are given in Table~\ref{tab:img}. Note that the core is not detected in 1997, while its intensity exceeds that of the northern hotspot in~2018--2019. \label{fig:x-1997-2018}} 
\end{figure}
\unskip

\begin{table}[H]
\caption{$S/X$-band VLBA image~parameters.\label{tab:img}}
\begin{adjustwidth}{-\extralength}{0cm}
\begin{tabularx}{\fulllength}{CCccccC}
\toprule
\textbf{Mean}	& \textbf{Frequency}  & \textbf{Peak} & \textbf{Lowest} & \multicolumn{2}{c}{\textbf{Restoring Beam} *} & \textbf{Figure} \\
\textbf{Epoch}	&  & \textbf{Brightness} & \textbf{Contour} & \textbf{FWHM} & \textbf{PA} &  \\
\textbf{[yr]}	& \textbf{[GHz]}  & \textbf{[mJy\,beam$^{-1}$]} & \textbf{[mJy\,beam$^{-1}$]} & \textbf{[mas $\times$ mas]} & \textbf{[$^{\circ}$]} & \\
\midrule
$1997.344$ & $2.27$  & $377$  & $\pm5$ & $16.2 \times 5.8$ & $11.8$ & Figure  \ref{fig:s-1997-2018} \\
$2019.124$ & $2.28$  & $356$  & $\pm2.5$ & $9.3 \times 3.3$ & $-9.5$ &  Figure \ref{fig:s-1997-2018} \\
$1997.344$ & $8.34$  & $89$  & $\pm4$ & $2.5 \times 1.0$ & $-3.4$ &  Figure \ref{fig:x-1997-2018} \\
$2019.124$ & $8.65$  & $72$  & $\pm2.5$ & $2.4 \times 0.9$ & $-9.1$ &  Figure \ref{fig:x-1997-2018} \\

\bottomrule
\end{tabularx}
\end{adjustwidth}
\noindent{\footnotesize{* The full width at half-maximum and major axis position angle (measured from north through east) of the elliptical Gaussian restoring beam}.}
\end{table}

\section{Discussion}
\label{sect:disc}
\unskip

\subsection{Astrometry}
 
It can be seen (Figure~\ref{fig:pos}) that the GIM-corrected $S$-band positions measured after 2014, while scattered, are generally consistent with the $S/X$ position obtained earlier in 1997 but~significantly offset from the other position solutions derived from the 2014--2020 observations. Furthermore, the~$S/X$ and $X$ (GIM-corrected) positions differ significantly from each other, which is caused by the fact that the source brightness peak positions at the $S$ and $X$ bands are clearly non-coincident (see Figures~\ref{fig:s-1997-2018} and \ref{fig:x-1997-2018}), and~thus the condition for forming the ionosphere-free linear combination (see Section~\ref{subsec:sx}) is not~met.  

At a given observing frequency band (in our case, $S$ or $X$), the~group delay for an interferometric baseline vector $\vec{B}$  with the source unit vector $\vec{s}_\mathrm{band}$ is presented as a sum of the geometric delay (including the relativistic effects), the~time delay caused by the neutral troposphere (dry and wet), and~the ionospheric delay. Herewith we assume that the delay for the troposphere is equal for both bands; thus it does not contribute to the additional $S-X$ bias. Therefore, a~simplified expression for the total delay composed of the geometric and ionospheric contributions can be written as follows:
\begin{equation}
  	\tau_\mathrm{band} = \frac{\vec{B} \cdot \vec{s}_\mathrm{band}}{c} + \frac{b}{f_\mathrm{band}^{2}}. 
  	\label{eq:delay_vec}
  \end{equation} 

Let us introduce $\vec{s}_{0}$ as the source vector independent of the frequency, and~consider decomposing the source vector at a particular frequency band as $\vec{s}_\mathrm{band} = \vec{s}_{0} + \Delta \vec{s}_\mathrm{band}$.
Then, for~the $X$ and $S$ bands, the~group delays can be written as
\begin{equation}
  	\tau_{X} = \frac{\vec{B}\cdot\vec{s}_{0}}{c}
  	+ \frac{ \vec{B}\cdot \Delta\vec{s}_X }{c} + \frac{b}{f_{X}^{2}},
  	\qquad
  	\tau_{S} = \frac{\vec{B}\cdot\vec{s}_{0}}{c}
  	+ \frac{ \vec{B}\cdot \Delta\vec{s}_S }{c} + \frac{b}{f_{S}^{2}}.
  	\label{eq:tau_sx}
  \end{equation}
Their difference is
\begin{equation}
  	\tau_{X} - \tau_{S}
  	= \frac{\vec{B}\cdot\Delta  \vec{s}_X }{c}-\frac{\vec{B}\cdot \Delta  \vec{s}_S }{c} 
  	+ b\ \left(\frac{1}{f_{X}^{2}} - \frac{1}{f_{S}^{2}}\right).
  	\label{eq:diff}
  \end{equation}  
The VLBI dual-band ($S/X$) calibration according to  Equation~(\ref{ion:1}) shifts the true value of $b$ to
\begin{equation}
  	b_{S/X} =
  	\frac{\tau_{X}-\tau_{S}}{\dfrac{1}{f_{X}^{2}}-\dfrac{1}{f_{S}^{2}}}
  	= b + \frac{\vec{B}\cdot (\Delta  \vec{s}_X -\Delta  \vec{s}_S)}
  	{c \left(\dfrac{1}{f_{X}^{2}}-\dfrac{1}{f_{S}^{2}} \right)}. 
  	\label{eq:best}
  \end{equation}
Equation~(\ref{eq:best}) shows that the estimate of parameter $b$ is unbiased if and only if $\Delta\vec{s}_X = \Delta\vec{s}_S$. The~conventional estimate of parameter $b$  is not correct if the radio source coordinates at the $S$ and $X$ bands are different. Consequently, the~standard $S/X$ calibration may result in a biased estimate of the source~position.

  The ionosphere-free delay $\tau_{0}$ is calculated using dual-band $S/X$ correction from Equation~(\ref{eq:tau0}) as
\begin{equation}
   	\tau_{0} = \tau_{X} - \frac{b_{S/X}}{f_{X}^{2}}  = \frac{\vec{B}\cdot\vec{s}_{0}}{c} +\frac{\vec{B}\cdot \Delta \vec{s}_X }{c}+  \frac{b}{f_{X}^{2}}
  	- \frac{b}{f_{X}^{2}}
  	- \frac{ \vec{B}\cdot (\Delta  \vec{s}_X -\Delta  \vec{s}_S) }{c f_{X}^{2}   \left(\dfrac{1}{f_{X}^{2}}-\dfrac{1}{f_{S}^{2}}\right)}
  	\label{eq:tau_corr}  
    \end{equation}  
By simplifying this,
\begin{align}   
    \tau_{0}
  	= \frac{\vec{B}\cdot\vec{s}_{X}}{c}  +
  	  	\frac{\vec{B} \cdot  ( \Delta  \vec{s}_X -\Delta  \vec{s}_S )}{c}
        \frac{f_{S}^{2}}{ f_{X}^{2}-f_{S}^{2}}.  
  	\label{eq:tau_corr2}
  \end{align}

$\Delta \vec{s}_{X}$ and $\Delta \vec{s}_{S}$ are the linear combinations of the angular offsets in right ascension and declination. By~linearising around a reference position $(\alpha_{0},\delta_{0})$, a~small
angular offset $(\Delta\alpha,\Delta\delta)$ changes the  source vector $\vec{s_0} = \bigl(\cos\alpha_0\cos\delta_0,\; \sin\alpha_0\cos\delta_0,\; \sin\delta_0\bigr)$  as
\begin{equation}
	 \Delta \vec{s}_\text{band} = \vec{s}_\text{band} -  \vec{s}_{0} \approx
	 \vec{e}_{\alpha}\,\Delta\alpha_\text{band}
	+ \vec{e}_{\delta}\,\Delta\delta_\text{band},
	\label{eq:linearise}
\end{equation}
with the basis vectors
\begin{equation}
	\vec{e}_{\alpha} =
	\begin{pmatrix}
		-\sin\alpha_{0}\cos\delta_{0} \\[2pt]
		\cos\alpha_{0}\cos\delta_{0} \\[2pt]
		0
	\end{pmatrix},
	\qquad
	\vec{e}_{\delta} =
	\begin{pmatrix}
		-\cos\alpha_{0}\sin\delta_{0} \\[2pt]
		-\sin\alpha_{0}\sin\delta_{0} \\[2pt]
		\cos\delta_{0}
	\end{pmatrix}.
	\label{eq:tangent}
\end{equation} 
The difference between the changes in the $X$ and $S$ bands is
\begin{equation} 
 	\Delta \vec{s}_X - \Delta \vec{s}_S = \vec{e}_{\alpha^*} (\Delta\alpha^*_X -\Delta\alpha_S^*) 
 	+ \vec{e}_{\delta} (\Delta\delta_X - \Delta\delta_S).
\end{equation}

Equation~(\ref{eq:linearise}) can be rewritten in terms of the corrected right ascension $\alpha^*=\alpha \cos \delta_0$~as
\begin{equation}
	 \Delta \vec{s}_\text{band}  \approx
	 \vec{e}_{\alpha^*}\,\Delta\alpha^*_\text{band}
	+ \vec{e}_{\delta}\,\Delta\delta_\text{band},
	\label{eq:linearise1}
\end{equation}
where $\vec{e}_{\alpha^*} = \vec{e}_{\alpha} / \cos \delta_0$ and $   \Delta\alpha^*_\text{band} = \Delta\alpha_\text{band} \cos \delta_0$. Hence,
\begin{equation}
  	   \Delta  \vec{s}_X= \vec{e}_{\alpha^*}\,\Delta\alpha_X^*
  	+ \vec{e}_{\delta}\,\Delta\delta_X, \quad
   \Delta  \vec{s}_S= \vec{e}_{\alpha^*}\,\Delta\alpha^*_S
  + \vec{e}_{\delta}\,\Delta\delta_S.
\end{equation} 

We can recover the angular offsets from Equation~(\ref{eq:tau_corr2}), where the position from $\tau_{0}$ corresponds to the $S/X$ position. The~derived $S/X$-band position 
satisfies
\begin{equation}
 \Delta\alpha_{S/X}^* = \Delta\alpha_{X}^* 
  	+ \frac{f_{S}^{2}}{f_{X}^{2}-f_{S}^{2}} (\Delta\alpha^*_X -\Delta\alpha_S^*), \quad
     \Delta\delta_{S/X} = \Delta\delta_{X} 
  	+ \frac{f_{S}^{2}}{f_{X}^{2}-f_{S}^{2}} (\Delta\delta_X -\Delta\delta_S).
	\label{eq:sxs_derived}
 \end{equation}
  
In contrast to dual-band ($S/X$) ionosphere calibration, the application of GIMs removes the true ionospheric delay and recovers the genuine centroid ($\Delta\alpha_{X}^*$, $\Delta\delta_{X}$).
 The coordinate offset is
\begin{equation}
   	\Delta\alpha_{S/X}^* - \Delta\alpha_{X}^* = K_{S}\,  (\Delta\alpha^*_X -\Delta\alpha_S^*) ,
   	\qquad
  	\Delta\delta_{S/X} - \Delta\delta_{X} = K_{S}\,(\Delta\delta_X -\Delta\delta_S),
   	\label{eq:coord_offset_vecX}
   \end{equation} 
where, for~the VLBA mean frequencies at the $S$ and $X$ bands,    $f_S$=2.27~GHz and $f_X$ = 8.67~GHz,
\begin{equation*}
  K_{S} = \frac{f_{S}^{2}}{f_{X}^{2}-f_{S}^{2}} \approx 0.075 .
     \end{equation*} 

Table~\ref{tab:gaia} presents the mean difference between the $S$- and $X$-band positions, which is approximately $-100$~mas in declination and $-27$~mas in corrected right ascension. This offset introduces a bias in the combined $S/X$ position of  
$-7.34$ mas in declination and 
$1.99$~mas in right ascension.  While the right ascension bias is consistent with the measured $S/X$ minus $X$ difference, ($1.89 \pm 0.15$)~mas (Table~\ref{tab:gaia}), falling within the $1.7\sigma$ confidence limits, the~declination shows a notable tension: the measured difference is ($-6.41\pm0.25$)~mas, resulting in a $\sim 3.7\sigma$ discrepancy from the expected bias. However, this difference is not critical, considering the $S$-band limitations: high susceptibility to radio frequency interference (RFI), lower signal-to-noise ratio compared to the $X$ band, and~the fact that we only modelled the first-order ionospheric~delay.

Geodetic/astrometric dual-band VLBI observations are traditionally carried out at the $S$ and $X$ bands. In~this scheme, the~$S$-band data are used for calculating the ionospheric delay. Since the $S$ band is heavily affected by RFI from artificial sources, those observations are made at  narrower bandwidths compared to the $X$ band. For~example, in~the VLBA observations used in this study, the~$S$-band frequency range was covered by four IFs, while the $X$ band is covered by twelve IFs. During~postprocessing, one $S$-band IF had to be discarded because of the high RFI level. Time delay dispersion is inversely proportional to the signal-to-noise ratio and the effective bandwidth. Thus, the~uncertainties are larger at the $S$ band, and~the positions obtained from $S$-band data with GIM corrections show larger scatter. In~addition, the~extended nature of the N hotspot region which contains the $S$-band brightness peak (Figure~\ref{fig:s-1997-2018}) may also contribute to the uncertainty of the position determination at this~band.

\begin{figure}[H]
  
\includegraphics[width=10.5cm]{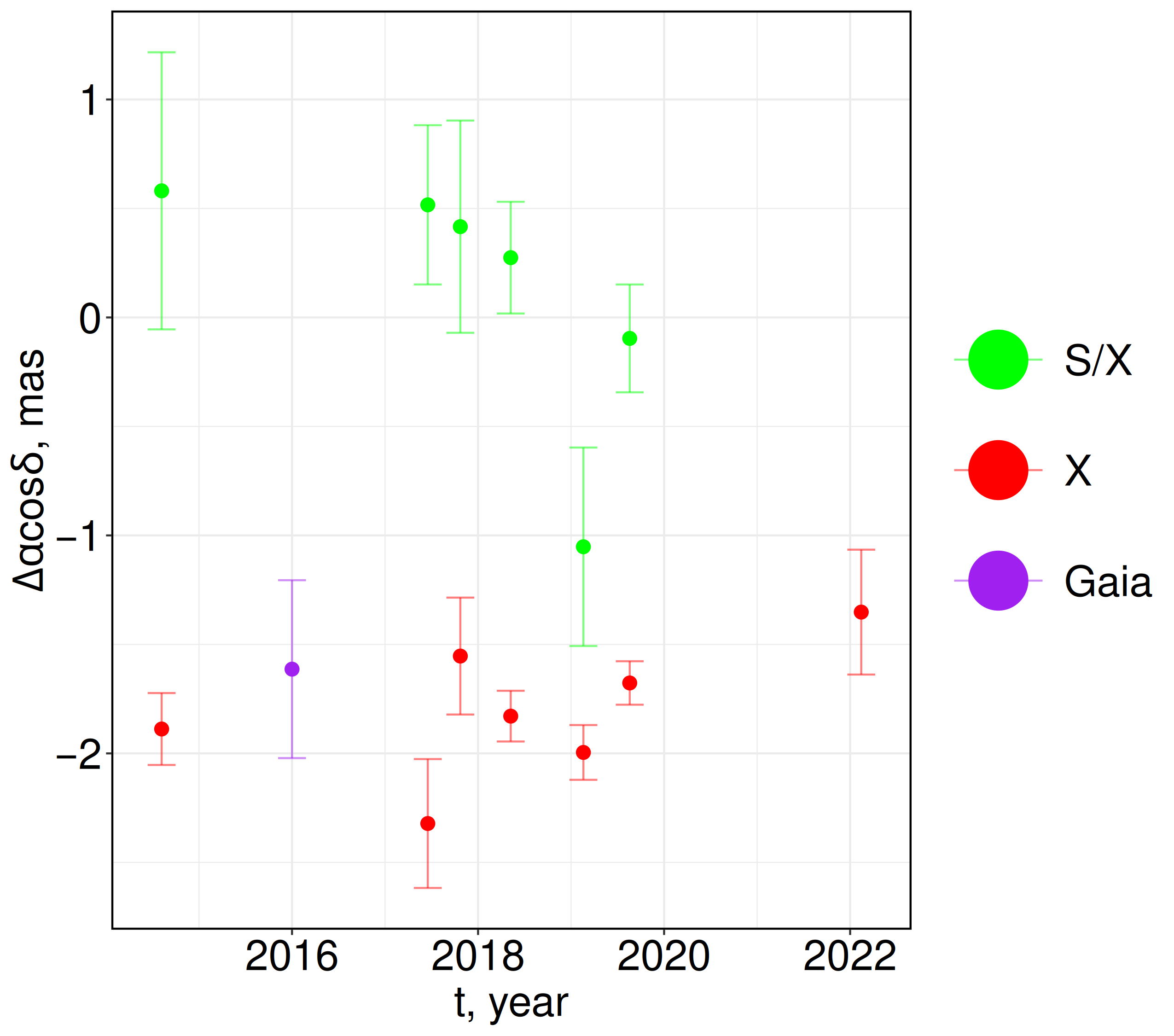}  
\includegraphics[width=10.5cm]{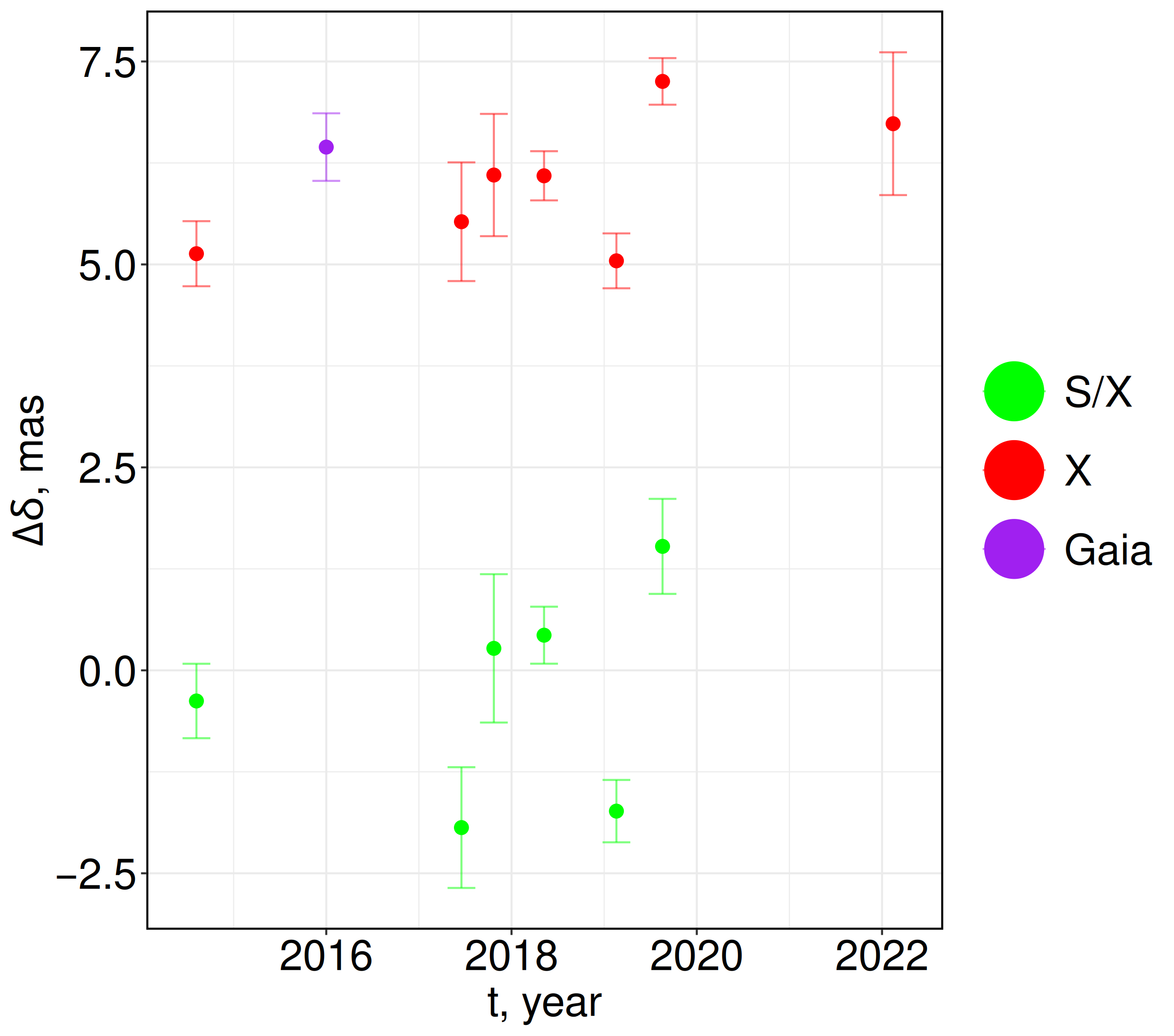}  
\caption{The post-2014 relative corrected right ascension (\textbf{top}) and declination (\textbf{bottom}) time series of 1524$-$136 at dual $S/X$ (green) and $X$ bands (red) in the radio, together with the \textit{Gaia} DR3 optical coordinates (purple), using a subset of data points displayed in Figure~\ref{fig:pos}.}\label{fig:gaia}
\end{figure}

As a result of the violation of the conditions of ionosphere-free linear combination (i.e., the~source positions at $S$ and $X$ bands are {very} different), the~comparison of various radio position solutions with the optical position from the \textit{Gaia} DR3 catalogue shows better agreement with the the $X$-band (GIM-corrected) position than with the $S/X$ position (Table~\ref{tab:gaia}, Figure~\ref{fig:gaia}), while a much larger offset (about $30$~mas in right ascension and $100$~mas in declination) is present with respect to the $S$-band coordinates (Figure~\ref{fig:pos}).

\unskip

\begin{table}[H]
    \caption{The difference between the \textit{Gaia} DR3 optical position and the post-2014 weighted mean $S/X$, $S$, and~$X$ radio positions. Frequency coefficient $K_S = 0.075$.
    }\label{tab:gaia}
    \begin{tabularx}{\textwidth}{cccccc}
    \toprule
      & \textit{\textbf{S/X}} \boldmath{$-$} \textit{\textbf{X}}     &  \textit{\textbf{S/X}} \boldmath{$-$} \textit{\textbf{Gaia}} &  \textit{\textbf{X}} \boldmath{$-$} \textit{\textbf{Gaia}} & \textit{\textbf{X}} \boldmath{$-$} \textit{\textbf{S}} & \boldmath{$K_S$} (\textit{\textbf{X}} \boldmath{$-$} \textit{\textbf{S}}) \\
           &  \textbf{[mas]} &   \textbf{[mas]} & \textbf{[mas]} & \textbf{[mas]} & \textbf{[mas]}\\
    \midrule        
   $\Delta\alpha$ *  & $1.89 \pm 0.15$ &  $1.70  \pm 0.43$ & $-0.19 \pm 0.41$ & $26.89 \pm  0.48$ & $1.99 \pm  0.04$ \\
       $\Delta\delta$ & $-6.41 \pm 0.25$ & $-6.78 \pm 0.46$ & $-0.38 \pm 0.44$ &
       $-99.17  \pm 1.11$ &  $-7.34  \pm 0.09$\\
       \bottomrule  
    \end{tabularx} 

{\noindent{\footnotesize{Here * denotes the corrected right ascension $\Delta\alpha\,\cos \delta_0$}.}
}
\end{table}

The fact that the $S$-band-only coordinates are consistent with the $S/X$ observation in 1997 suggests that the $S$- and $X$-band brightness peaks coincided at that epoch. This is supported by the $S$- and $X$-band VLBI images from that time (Figures~\ref{fig:s-1997-2018} and \ref{fig:x-1997-2018}). However, the~individual $S$- and $X$-band GIM-corrected positions could not be derived for 1997, as~the suitable GNSS-based daily GIMs are only available from 1998~onwards. 

{The phenomenon discussed here is also relevant for the geodetic VLBI observations performed with the next-generation broadband  VLBI Global Observing System (VGOS,~\cite{2018RaSc...53.1269N}). Extended source structure, which also leads to significant discrepancies between coordinates obtained at different frequencies, is among the most serious limitations of VGOS astrometry and geodesy~\cite{2021JGRB..12621238X}. The~source 1524$-$136 serves as an extreme example of this problem. While the broad frequency coverage improves group-delay precision compared to traditional dual-band ($S/X$) observations, it can be even more severely affected by the frequency-dependent differences in the source brightness distribution. 
Correcting source structure effects based on VLBI images  is a promising way to mitigate systematics in VGOS measurements~\cite{2021JGRB..12621238X}.}

\subsection{Imaging}

The snapshot VLBI images of 1524$-$136 presented here (\cref{fig:c-1996,fig:s-1997-2018,fig:x-1997-2018}) are generally consistent with those published in the literature~\cite{2002A&A...389L..15M,2002A&A...389...58M}. They recover the core and northern jet region but are not sensitive to the faint and diffuse emission on the counter-jet side to the southwest. From~a VLBI astrometric point of view, however, the~counter-jet is not relevant because of its relative insignificance within the overall source brightness~distribution. 

At the lowest frequencies studied here ($S$ band) and available in the literature ($L$ band, $1.7$~GHz~\cite{2002A&A...389L..15M}), the~component N, known to be a terminal hotspot at the end of the northern jet, dominates the brightness distribution at all times (Figure~\ref{fig:s-1997-2018}). While this was also true at higher frequencies in the 1990s ($C$ band in Figure~\ref{fig:c-1996} and~\cite{2002A&A...389...58M} and~$X$ band in Figure~\ref{fig:x-1997-2018} and~\cite{2002A&A...389L..15M}), the~core component C became the brightest feature by the time, after~a long gap, new VLBI observations of 1524$-$136 were  performed in the 2010s (see the 2018--2019 image in Figure~\ref{fig:x-1997-2018}). This brightening of the core relative to the jet hotspot caused a $\sim100$-mas shift in the $X$-band astrometric position, while the $S$-band position defined by the jet component remained roughly the same throughout more than two decades covered by the astrometric VLBI measurements (Figure~\ref{fig:pos}). 

At higher frequencies, the~steep power-law radio spectrum causes N {to} appear fainter, and~the increased angular resolution of the same interferometer array starts to resolve  its extended emission. On~the other hand, its compactness and flat spectrum makes the core (C) the dominant feature at the $X$ band in the~2010s.

{While there are no VLBI imaging data from 1997 to 2014, total flux density monitoring data with ATCA at multiple frequencies between 1996 and 2000~\cite{2003PASJ...55..351T} are available. In~the case of total flux density variability, it is reasonable to assume that the compact core and not the extended structure is responsible for it. Therefore,~major core brightening could in principle be inferred from an increased total flux density. However, the~monitoring data show little variability in 1524$-$136, which had $0.75$~Jy mean flux density at the $X$ band, with~a small variability index $0.03$ \cite{2003PASJ...55..351T}. This suggests that  VLBI core brightening might have occurred after 2000.}

{More recent multi-band flux density monitoring data are also available from between 2005 and 2023, in~the ATCA Calibrator Database (\url{https://www.narrabri.atnf.csiro.au/calibrators/calibrator_database_viewcal?source=1524$-$136}, accessed on 23 October 2026). Significant variability is not seen there either, and~the comparison with the older data is not straightforward because of the different frequency coverages. Moreover, if~the contribution of extended radio lobes and hotspots to the total flux density is considerable and relatively stable in time, like probably in the case of 1524$-$136, the~total flux density may exhibit only modest variability even when the compact core varies substantially. Nevertheless, a~core flux density increase of $\sim 0.1$~Jy which would explain  core brightening appears consistent with the total flux density measurements, tentatively narrowing down the time range for  core brightening to between 2000 and 2005.}       

\subsection{Possible Physical Explanations of Core~Brightening}

{Jetted radio AGN are known to be variable on a wide range of time scales, including years--decades. This is not restricted to blazars with jets closely aligned to our line of sight but also observed in other AGN (e.g.,~\cite{2004AJ....127.2565D,2020ApJ...905...74N,2026ApJ..1003...46K}). Variability can naturally be explained by changes in  compact synchrotron-emitting regions and the~emergence and evolution of jet~components.} 

Since 1524$-$136 exhibits an extended radio structure characteristic of past jet activity, and~jetted AGN are known to be recurrent~\cite{2024Galax..12...11M}, it is intriguing to speculate about the possibility of whether we can catch a restarting jet in the act in the core of this quasar. In~this scenario, the~northern hotspot would represent a previous cycle of activity, while the brightening central component signals the emergence of a new compact jet. In~some models of black hole spin evolution~\cite{2019ApJ...871..259G,2026Galax..14....2G}, this could correspond to a transition from retrograde to prograde accretion. However, for~the new jet to emerge so quickly, i.e.,~before the relic emission fades away, a high (near-Eddington) accretion rate is required. Such a rapid evolution would be more likely in a relatively low-density large-scale environment. Rapidly accreting radio AGN have been associated with poor environments~\cite{2020ApJ...905..156G}. Concerning 1524$-$136 in particular, the deep infrared imaging of the field surrounding the source~\cite{2003MNRAS.343....1B} revealed a galaxy overdensity. While a moderate excess of red galaxies was found on scales out to about $1$ Mpc, there is no sharp central peak of the galaxy distribution. This suggests that the source does not reside in a cluster environment but perhaps in a forming cluster. From~the available observations, it is not easy to distinguish between the usual variability in  jet activity and the restarted jet scenario, but~the former seems more likely. In~any case, the~astrophysical scenario does not affect  the observed astrometric behaviour of 1524$-$136.

\section{Summary and~Conclusions}
\label{sect:concl}

In this paper, we analysed the coordinates and images of the extragalactic source 1524$-$136. In~its traditional ionosphere-corrected $S/X$ VLBI coordinate time series, the~first astrometric observation (1997) deviates from the group of later observations (after 2014) by more than 100 mas. The~1997 position corresponds to the bright hotspot in the northern jet, while the post-2014 observations are dominated by the VLBI core. We cannot precisely determine when the core brightened between 1997 and 2014 {(or perhaps between 2000 and 2005, as~inferred from total flux density monitoring data)}, as~there are no VLBI observations of this source available from that period. This points to the importance of regularly monitoring VLBI structures and the positions of ICRF~sources. 

In addition to the combined $S/X$ coordinates, we obtained the source position individually at both the $S$ and $X$ bands by applying ionospheric corrections using GNSS-based GIMs for VLBI observations from 2014 to 2022. We found that the $S$- and $X$-band positions refer to different widely displaced structural components. The~$S$-band position is dominated by the northern jet throughout the entire time covered by VLBI observations, whereas the $X$-band position refers to the compact VLBI core at post-2014 epochs. As~a result of the inconsistency between the $S$- and $X$-band brightness peaks, the~standard ionospheric calibration does not work properly and introduces a bias in the combined $S/X$ position. The~difference between $X$-band-only (GIM calibration) and the $S/X$ position is about $6.9$~mas. On~the other hand, there is no significant difference between the $X$-band radio and the \textit{Gaia} DR3 optical position, while a significant offset is measured between the \textit{Gaia} and  $S/X$ positions. This suggests that the use of external ionospheric calibration from GNSS-based GIMs makes the $X$-band-only radio source position solutions less susceptible in similar cases of structural mismatch between the $S$ and $X$ bands than the traditional $S/X$~solutions. 

In the context of preparing for the next ICRF release, for~sources exhibiting a significant difference between their $S$- and $X$-band positions, ionospheric corrections should account for this effect. A~problem similar to that of 1524$-$136 may arise for sources with multiple components separated by large angular distances (several 10s of mas or more), as~well as for sources showing large time variations in their $S/X$ coordinates, as~was demonstrated recently for the radio source 1328+254~\cite{Urunova2025}.

\authorcontributions{Conceptualisation, A.O. and S.F.; methodology, A.O., O.T., A.M. and S.F.; software, O.T. and A.M.; formal analysis, A.O. and P.G.; validation, A.M.; writing---original draft preparation, A.O. and S.F.; writing---review and editing, O.T., A.M., and P.G.; visualisation, A.O. and S.F.; supervision, O.T. and S.F. All authors have read and agreed to the published version of the~manuscript.}  

\funding{This research received no external~funding.}

\dataavailability{The calibrated VLBI visibility data used for imaging in this article are available in the Astrogeo Center database (\url{https://astrogeo.org/cgi-bin/imdb_get_source.csh?source=J1526-1351}, accessed on 5 October 2026). 
The astrometric IVS datasets were derived from sources in the public domain (\url{https://ivscc.gsfc.nasa.gov/products-data/data.html}, accessed on 5 October 2026).
The V515E project data are available in the LBA archive (\url{https://storage.pawsey.org.au/public/m/VLBI/Archive/LBA/v515/v515e}, accessed on 5 October 2026).
The calculation of the ionospheric contribution was performed using GIMs produced by the CODE Analysis Center, which are available at the Astronomical Institute of University of Bern (\url{https://code.aiub.unibe.ch/s3_script/aiub_s3_bucket_listing.php?path=CODE}, accessed on 5 October 2026).}

\acknowledgments{
{We are grateful for the constructive and insightful comments and suggestions received from three anonymous~reviewers.}
We thank Jamie McCallum for postprocessing the V515E correlated data.
We are indebted to the International VLBI Service for Geodesy and Astrometry~\cite{2017JGeod..91..711N} for providing the results of the VLBI observations.
We are grateful to the CODE Analysis Center~\cite{Schaer1996} in the Astronomical Institute of University of Bern for providing the global ionosphere map data.
Quasar VLBI Network radio telescopes have been participating in the V515 project as ad hoc VLBI stations. Svetloe and Zelenchukskaya radio telescopes are operated by the Scientific Equipment Sharing Center of the Quasar VLBI Network.
The Australia Telescope Compact Array, Murriyang, CSIRO’s Parkes radio telescope, the~Mopra telescope, and the~Long Baseline Array are parts of the Australia Telescope National Facility~(\url{https://ror.org/05qajvd42}, accessed on 5 October 2026) which is funded by the Australian Government for operation as a National Facility managed by CSIRO.
Host country access to Tidbinbilla antennas is arranged by the ATNF.
This work was supported by resources provided by the Pawsey Supercomputing Research Centre with funding from the Australian Government and the Government of Western Australia. The National Radio Astronomy Observatory is a facility of the National Science Foundation operated under cooperative agreement by Associated Universities, Inc. 
The authors acknowledge the use of the Astrogeo Center database maintained by Leonid Petrov (\url{https://doi.org/10.25966/kyy8-yp57}, accessed on 5 October 2026).
}

\conflictsofinterest{The authors declare no conflicts of~interest.} 

\abbreviations{Abbreviations}{
The following abbreviations are used in this manuscript:\vspace{-12pt}
\\

\noindent 
\begin{longtable}[l]{@{}ll} 

AGN & Active Galactic Nuclei\\
ATCA & Australia Telescope Compact Array \\
ATNF & Australia Telescope National Facility \\
CODE & Center for Orbit Determination in Europe \\
CSIRO & Commonwealth Scientific and Industrial Research Organisation \\
CSS & Compact Steep Spectrum (quasar)\\
DR & Data Release\\
FWHM & Full Width at Half-Maximum \\
GIM & Global Ionosphere Map \\
GNSS & Global Navigation Satellite System\\
HOPS & Haystack Observatory Postprocessing System\\
HV & Horizontal and Vertical (linear polarisation) \\
ICRF & International Celestial Reference Frame\\
IF & Intermediate Frequency Channel\\
IVS & International VLBI Service for Geodesy and Astrometry\\
LBA & Long Baseline Array\\
mas & Milliarcsecond \\
MJD & Modified Julian Date\\
PA & Position Angle\\
RCP & Right Circular Polarisation \\
RFI & Radio Frequency Interference\\
sTEC & Slant Total Electron Content\\
TEC & Total Electron Content\\
TECU & Total Electron Content Unit\\
{VGOS} & {VLBI Global Observing System} \\
VLA & Karl G. Jansky Very Large Array\\
VLBI & Very Long Baseline Interferometry \\
VLBA & Very Long Baseline Array \\
\end{longtable}
}

\begin{adjustwidth}{-\extralength}{0cm}
\printendnotes[custom] 

\reftitle{References}

\PublishersNote{}
\end{adjustwidth}

\begin{thebibliography}{999}

\bibitem[{Blandford} et~al.(2019){Blandford}, {Meier}, and
  {Readhead}]{2019ARA&A..57..467B}
{Blandford}, R.; {Meier}, D.; {Readhead}, A.
\newblock {Relativistic Jets from Active Galactic Nuclei}.
\newblock {\em \araa} {\bf 2019}, {\em 57},~467--509. 
\newblock {\url{https://doi.org/10.1146/annurev-astro-081817-051948}}.

\bibitem[{Fabian}(2012)]{2012ARA&A..50..455F}
{Fabian}, A.C.
\newblock {Observational Evidence of Active Galactic Nuclei Feedback}.
\newblock {\em \araa} {\bf 2012}, {\em 50},~455--489.
\newblock {\url{https://doi.org/10.1146/annurev-astro-081811-125521}}.

\bibitem[{Mukherjee}(2025)]{2025Galax..13..102M}
{Mukherjee}, D.
\newblock {Jet Feedback on kpc Scales: A Review}.
\newblock {\em Galaxies} {\bf 2025}, {\em 13},~102.
\newblock {\url{https://doi.org/10.3390/galaxies13050102}}.

\bibitem[{Ghodsi Yengejeh} et~al.(2026){Ghodsi Yengejeh}, {Koryukova},
  {Gurvits}, {Frey}, {Pushkarev}, {Plavin}, {Kellermann}, and
  {Kov{\'a}cs}]{2026A&A...710A.133G}
{Ghodsi Yengejeh}, M.; {Koryukova}, T.A.; {Gurvits}, L.I.; {Frey}, S.;
  {Pushkarev}, A.B.; {Plavin}, A.V.; {Kellermann}, K.I.; {Kov{\'a}cs}, A.
\newblock {Revisiting the angular size--redshift cosmological test with
  milliarcsecond radio structures in active galactic nuclei}.
\newblock {\em \aap} {\bf 2026}, {\em 710},~A133.
\newblock {\url{https://doi.org/10.1051/0004-6361/202557047}}.

\bibitem[{Hada}(2020)]{2019Galax...8....1H}
{Hada}, K.
\newblock {Relativistic Jets from AGN Viewed at Highest Angular Resolution}.
\newblock {\em Galaxies} {\bf 2020}, {\em 8},~1.
\newblock {\url{https://doi.org/10.3390/galaxies8010001}}.

\bibitem[{Sovers} et~al.(1998){Sovers}, {Fanselow}, and
  {Jacobs}]{1998RvMP...70.1393S}
{Sovers}, O.J.; {Fanselow}, J.L.; {Jacobs}, C.S.
\newblock {Astrometry and geodesy with radio interferometry: Experiments,
  models, results}.
\newblock {\em Rev. Mod. Phys.} {\bf 1998}, {\em 70},~1393--1454.
\newblock {\url{https://doi.org/10.1103/RevModPhys.70.1393}}.

\bibitem[{de Witt} et~al.(2022){de Witt}, {Charlot}, {Gordon}, and
  {Jacobs}]{2022Univ....8..374D}
{de Witt}, A.; {Charlot}, P.; {Gordon}, D.; {Jacobs}, C.S.
\newblock {Overview and Status of the International Celestial Reference Frame
  as Realized by VLBI}.
\newblock {\em Universe} {\bf 2022}, {\em 8},~374.
\newblock {\url{https://doi.org/10.3390/universe8070374}}.

\bibitem[{Sokolovsky} et~al.(2011){Sokolovsky}, {Kovalev}, {Pushkarev}, and
  {Lobanov}]{2011A&A...532A..38S}
{Sokolovsky}, K.V.; {Kovalev}, Y.Y.; {Pushkarev}, A.B.; {Lobanov}, A.P.
\newblock {A VLBA survey of the core shift effect in AGN jets. I. Evidence of
  dominating synchrotron opacity}.
\newblock {\em \aap} {\bf 2011}, {\em 532},~A38.
\newblock {\url{https://doi.org/10.1051/0004-6361/201016072}}.

\bibitem[{Mo{\'o}r} et~al.(2011){Mo{\'o}r}, {Frey}, {Lambert}, {Titov}, and
  {Bakos}]{2011AJ....141..178M}
{Mo{\'o}r}, A.; {Frey}, S.; {Lambert}, S.B.; {Titov}, O.A.; {Bakos}, J.
\newblock {On the Connection of the Apparent Proper Motion and the VLBI
  Structure of Compact Radio Sources}.
\newblock {\em \aj} {\bf 2011}, {\em 141},~178.
\newblock {\url{https://doi.org/10.1088/0004-6256/141/6/178}}.

\bibitem[{Titov} et~al.(2022){Titov}, {Frey}, {Melnikov}, {Lambert}, {Shu},
  {Xia}, {Gonz{\'a}lez}, {Tercero}, {Gulayev}, {Weston}, and
  {Natusch}]{2022MNRAS.512..874T}
{Titov}, O.; {Frey}, S.; {Melnikov}, A.; {Lambert}, S.; {Shu}, F.; {Xia}, B.;
  {Gonz{\'a}lez}, J.; {Tercero}, B.; {Gulayev}, S.; {Weston}, S.;  et~al.
\newblock {Unprecedented change in the position of four radio sources}.
\newblock {\em \mnras} {\bf 2022}, {\em 512},~874--883.
\newblock {\url{https://doi.org/10.1093/mnras/stac038}}.

\bibitem[Osetrova et~al.(2024)Osetrova, Titov, and Melnikov]{Osetrova2024}
Osetrova, A.A.; Titov, O.A.; Melnikov, A.E.
\newblock Extragalactic Radio Sources with Large Changes in Their Apparent
  Coordinates.
\newblock {\em Astron. Lett.} {\bf 2024}, {\em 50},~657--670.
\newblock {\url{https://doi.org/10.1134/S1063773725700045}}.

\bibitem[{Gaia Collaboration} et~al.(2016){Gaia Collaboration}, {Prusti}, {de
  Bruijne}, {Brown}, {Vallenari}, {Babusiaux}, {Bailer-Jones}, {Bastian},
  {Biermann}, {Evans}, {Eyer}, {Jansen}, {Jordi}, {Klioner}, {Lammers},
  {Lindegren}, {Luri}, {Mignard}, {Milligan}, {Panem}, {Poinsignon},
  {Pourbaix}, {Randich}, {Sarri}, {Sartoretti}, {Siddiqui}, {Soubiran},
  {Valette}, {van Leeuwen}, {Walton}, {Aerts}, {Arenou}, {Cropper}, {Drimmel},
  {H{\o}g}, {Katz}, {Lattanzi}, {O'Mullane}, {Grebel}, {Holland}, {Huc},
  {Passot}, {Bramante}, {Cacciari}, {Casta{\~n}eda}, {Chaoul}, {Cheek}, {De
  Angeli}, {Fabricius}, {Guerra}, {Hern{\'a}ndez}, {Jean-Antoine-Piccolo},
  {Masana}, {Messineo}, {Mowlavi}, {Nienartowicz}, {Ord{\'o}{\~n}ez-Blanco},
  {Panuzzo}, {Portell}, {Richards}, {Riello}, {Seabroke}, {Tanga},
  {Th{\'e}venin}, {Torra}, {Els}, {Gracia-Abril}, {Comoretto},
  {Garcia-Reinaldos}, {Lock}, {Mercier}, {Altmann}, {Andrae}, {Astraatmadja},
  {Bellas-Velidis}, {Benson}, {Berthier}, {Blomme}, {Busso}, {Carry},
  {Cellino}, {Clementini}, {Cowell}, {Creevey}, {Cuypers}, {Davidson}, {De
  Ridder}, {de Torres}, {Delchambre}, {Dell'Oro}, {Ducourant}, {Fr{\'e}mat},
  {Garc{\'\i}a-Torres}, {Gosset}, {Halbwachs}, {Hambly}, {Harrison}, {Hauser},
  {Hestroffer}, {Hodgkin}, {Huckle}, {Hutton}, {Jasniewicz}, {Jordan},
  {Kontizas}, {Korn}, {Lanzafame}, {Manteiga}, {Moitinho}, {Muinonen},
  {Osinde}, {Pancino}, {Pauwels}, {Petit}, {Recio-Blanco}, {Robin}, {Sarro},
  {Siopis}, {Smith}, {Smith}, {Sozzetti}, {Thuillot}, {van Reeven}, {Viala},
  {Abbas}, {Abreu Aramburu}, {Accart}, {Aguado}, {Allan}, {Allasia},
  {Altavilla}, {{\'A}lvarez}, {Alves}, {Anderson}, {Andrei}, {Anglada Varela},
  {Antiche}, {Antoja}, {Ant{\'o}n}, {Arcay}, {Atzei}, {Ayache}, {Bach},
  {Baker}, {Balaguer-N{\'u}{\~n}ez}, {Barache}, {Barata}, {Barbier}, {Barblan},
  {Baroni}, {Barrado y Navascu{\'e}s}, {Barros}, {Barstow}, {Becciani},
  {Bellazzini}, {Bellei}, {Bello Garc{\'\i}a}, {Belokurov}, {Bendjoya},
  {Berihuete}, {Bianchi}, {Bienaym{\'e}}, {Billebaud}, {Blagorodnova},
  {Blanco-Cuaresma}, {Boch}, {Bombrun}, {Borrachero}, {Bouquillon}, {Bourda},
  {Bouy}, {Bragaglia}, {Breddels}, {Brouillet}, {Br{\"u}semeister},
  {Bucciarelli}, {Budnik}, {Burgess}, {Burgon}, {Burlacu}, {Busonero}, {Buzzi},
  {Caffau}, {Cambras}, {Campbell}, {Cancelliere}, {Cantat-Gaudin}, {Carlucci},
  {Carrasco}, {Castellani}, {Charlot}, {Charnas}, {Charvet}, {Chassat},
  {Chiavassa}, {Clotet}, {Cocozza}, {Collins}, {Collins}, and
  {Costigan}]{2016A&A...595A...1G}
{Gaia Collaboration}.; {Prusti}, T.; {de Bruijne}, J.H.J.; {Brown}, A.G.A.;
  {Vallenari}, A.; {Babusiaux}, C.; {Bailer-Jones}, C.A.L.; {Bastian}, U.;
  {Biermann}, M.; {Evans}, D.W.;  et~al.
\newblock {The Gaia mission}.
\newblock {\em \aap} {\bf 2016}, {\em 595},~A1.
\newblock {\url{https://doi.org/10.1051/0004-6361/201629272}}.

\bibitem[{Orosz} and {Frey}(2013)]{2013A&A...553A..13O}
{Orosz}, G.; {Frey}, S.
\newblock {Optical-radio positional offsets for active galactic nuclei}.
\newblock {\em \aap} {\bf 2013}, {\em 553},~A13.
\newblock {\url{https://doi.org/10.1051/0004-6361/201321279}}.

\bibitem[{Mignard} et~al.(2016){Mignard}, {Klioner}, {Lindegren}, {Bastian},
  {Bombrun}, {Hern{\'a}ndez}, {Hobbs}, {Lammers}, {Michalik}, {Ramos-Lerate},
  {Biermann}, {Butkevich}, {Comoretto}, {Joliet}, {Holl}, {Hutton}, {Parsons},
  {Steidelm{\"u}ller}, {Andrei}, {Bourda}, and {Charlot}]{2016A&A...595A...5M}
{Mignard}, F.; {Klioner}, S.; {Lindegren}, L.; {Bastian}, U.; {Bombrun}, A.;
  {Hern{\'a}ndez}, J.; {Hobbs}, D.; {Lammers}, U.; {Michalik}, D.;
  {Ramos-Lerate}, M.;  et~al.
\newblock {Gaia Data Release 1. Reference frame and optical properties of ICRF
  sources}.
\newblock {\em \aap} {\bf 2016}, {\em 595},~A5.
\newblock {\url{https://doi.org/10.1051/0004-6361/201629534}}.

\bibitem[{Petrov} and {Kovalev}(2017)]{2017MNRAS.467L..71P}
{Petrov}, L.; {Kovalev}, Y.Y.
\newblock {On significance of VLBI/Gaia position offsets}.
\newblock {\em \mnras} {\bf 2017}, {\em 467},~L71--L75.
\newblock {\url{https://doi.org/10.1093/mnrasl/slx001}}.

\bibitem[{Kovalev} et~al.(2017){Kovalev}, {Petrov}, and
  {Plavin}]{2017A&A...598L...1K}
{Kovalev}, Y.Y.; {Petrov}, L.; {Plavin}, A.V.
\newblock {VLBI-Gaia offsets favor parsec-scale jet direction in active
  galactic nuclei}.
\newblock {\em \aap} {\bf 2017}, {\em 598},~L1,
  \href{http://arxiv.org/abs/1611.02632}{{\normalfont
  [arXiv:astro-ph.GA/1611.02632]}}.
\newblock {\url{https://doi.org/10.1051/0004-6361/201630031}}.

\bibitem[{Plavin} et~al.(2019){Plavin}, {Kovalev}, and
  {Petrov}]{2019ApJ...871..143P}
{Plavin}, A.V.; {Kovalev}, Y.Y.; {Petrov}, L.Y.
\newblock {Dissecting the AGN Disk-Jet System with Joint VLBI-Gaia Analysis}.
\newblock {\em \apj} {\bf 2019}, {\em 871},~143.
\newblock {\url{https://doi.org/10.3847/1538-4357/aaf650}}.

\bibitem[{Lambert} et~al.(2024){Lambert}, {Sol}, and
  {Pierron}]{2024A&A...684A.202L}
{Lambert}, S.; {Sol}, H.; {Pierron}, A.
\newblock {Identification of the optical emission detected by Gaia with radio
  structures in parsec-scale active galactic nucleus jets}.
\newblock {\em \aap} {\bf 2024}, {\em 684},~A202.
\newblock {\url{https://doi.org/10.1051/0004-6361/202347210}}.

\bibitem[{Popkov} et~al.(2025){Popkov}, {Kovalev}, {Plavin}, {Petrov}, and
  {Pashchenko}]{2025MNRAS.543..479P}
{Popkov}, A.V.; {Kovalev}, Y.Y.; {Plavin}, A.V.; {Petrov}, L.Y.; {Pashchenko},
  I.N.
\newblock {Dim cores of radio-bright AGN jets: VLBI and Gaia astrometry
  pinpoint different parsec-scale features}.
\newblock {\em \mnras} {\bf 2025}, {\em 543},~479--495.
\newblock {\url{https://doi.org/10.1093/mnras/staf1480}}.

\bibitem[{Hewitt} and {Burbidge}(1993)]{1993ApJS...87..451H}
{Hewitt}, A.; {Burbidge}, G.
\newblock {A Revised and Updated Catalog of Quasi-stellar Objects}.
\newblock {\em \apjs} {\bf 1993}, {\em 87},~451--947.
\newblock {\url{https://doi.org/10.1086/191811}}.

\bibitem[{Bolton} and {Ekers}(1966)]{1966AuJPh..19..559B}
{Bolton}, J.G.; {Ekers}, J.
\newblock {Identification of strong extragalactic radio sources in the
  declination zone 0{\textdegree} to $-$20{\textdegree}}.
\newblock {\em Aust. J. Phys.} {\bf 1966}, {\em 19},~559.
\newblock {\url{https://doi.org/10.1071/PH660559}}.

\bibitem[{Mantovani} et~al.(1992){Mantovani}, {Junor}, {Fanti}, {Padrielli},
  {Browne}, and {Muxlow}]{1992MNRAS.257..353M}
{Mantovani}, F.; {Junor}, W.; {Fanti}, R.; {Padrielli}, L.; {Browne}, I.W.A.;
  {Muxlow}, T.W.B.
\newblock {Obervations at arcsecond resolution of steep-spectrum sources which
  vary at low frequencies.}
\newblock {\em \mnras} {\bf 1992}, {\em 257},~353--367.
\newblock {\url{https://doi.org/10.1093/mnras/257.2.353}}.

\bibitem[{Steppe} et~al.(1995){Steppe}, {Jeyakumar}, {Saikia}, and
  {Salter}]{1995A&AS..113..409S}
{Steppe}, H.; {Jeyakumar}, S.; {Saikia}, D.J.; {Salter}, C.J.
\newblock {Millimeter-wavelength observations of compact steep-spectrum
  sources.}
\newblock {\em \aaps} {\bf 1995}, {\em 113},~409--418.

\bibitem[{O'Dea} and {Saikia}(2021)]{2021A&ARv..29....3O}
{O'Dea}, C.P.; {Saikia}, D.J.
\newblock {Compact steep-spectrum and peaked-spectrum radio sources}.
\newblock {\em \aapr} {\bf 2021}, {\em 29},~3.
\newblock {\url{https://doi.org/10.1007/s00159-021-00131-w}}.

\bibitem[{Wright}(2006)]{2006PASP..118.1711W}
{Wright}, E.L.
\newblock {A Cosmology Calculator for the World Wide Web}.
\newblock {\em \pasp} {\bf 2006}, {\em 118},~1711--1715.
\newblock {\url{https://doi.org/10.1086/510102}}.

\bibitem[{Mantovani} et~al.(1994){Mantovani}, {Junor}, {Fanti}, {Padrielli},
  and {Saikia}]{1994A&A...292...59M}
{Mantovani}, F.; {Junor}, W.; {Fanti}, R.; {Padrielli}, L.; {Saikia}, D.J.
\newblock {Gaseous cocoons around compact steep spectrum sources.}
\newblock {\em \aap} {\bf 1994}, {\em 292},~59--66.

\bibitem[{Reid} et~al.(1999){Reid}, {Kronberg}, and
  {Perley}]{1999ApJS..124..285R}
{Reid}, R.I.; {Kronberg}, P.P.; {Perley}, R.A.
\newblock {VLA Images at 5 GHz of 212 Southern Extragalactic Objects}.
\newblock {\em \apjs} {\bf 1999}, {\em 124},~285--381.
\newblock {\url{https://doi.org/10.1086/313259}}.

\bibitem[{Kunert-Bajraszewska} and {Marecki}(2007)]{2007A&A...469..437K}
{Kunert-Bajraszewska}, M.; {Marecki}, A.
\newblock {FIRST-based survey of compact steep spectrum sources. V.
  Milliarcsecond-scale morphology of CSS objects}.
\newblock {\em \aap} {\bf 2007}, {\em 469},~437--450,
  \href{http://arxiv.org/abs/0704.0351}{{\normalfont
  [arXiv:astro-ph/0704.0351]}}.
\newblock {\url{https://doi.org/10.1051/0004-6361:20066364}}.

\bibitem[{Dallacasa} et~al.(2013){Dallacasa}, {Orienti}, {Fanti}, {Fanti}, and
  {Stanghellini}]{2013MNRAS.433..147D}
{Dallacasa}, D.; {Orienti}, M.; {Fanti}, C.; {Fanti}, R.; {Stanghellini}, C.
\newblock {A sample of small-sized compact steep-spectrum radio sources: VLBI
  images and VLA polarization at 5 GHz}.
\newblock {\em \mnras} {\bf 2013}, {\em 433},~147--161.
\newblock {\url{https://doi.org/10.1093/mnras/stt710}}.

\bibitem[{Fomalont} et~al.(2000){Fomalont}, {Frey}, {Paragi}, {Gurvits},
  {Scott}, {Taylor}, {Edwards}, and {Hirabayashi}]{2000ApJS..131...95F}
{Fomalont}, E.B.; {Frey}, S.; {Paragi}, Z.; {Gurvits}, L.I.; {Scott}, W.K.;
  {Taylor}, A.R.; {Edwards}, P.G.; {Hirabayashi}, H.
\newblock {The VSOP 5 GHz Continuum Survey: The Prelaunch VLBA Observations}.
\newblock {\em \apjs} {\bf 2000}, {\em 131},~95--183.
\newblock {\url{https://doi.org/10.1086/317368}}.

\bibitem[{Mantovani} et~al.(2002{\natexlab{a}}){Mantovani}, {Saikia}, {Bondi},
  {Junor}, {Salter}, and {Ricci}]{2002A&A...389L..15M}
{Mantovani}, F.; {Saikia}, D.J.; {Bondi}, M.; {Junor}, W.; {Salter}, C.J.;
  {Ricci}, R.
\newblock {B1524$-$136: A CSS quasar with two-sided radio jets}.
\newblock {\em \aap} {\bf 2002}, {\em 389},~L15--L19.
\newblock {\url{https://doi.org/10.1051/0004-6361:20020697}}.

\bibitem[{Mantovani} et~al.(2002{\natexlab{b}}){Mantovani}, {Junor}, {Ricci},
  {Saikia}, {Salter}, and {Bondi}]{2002A&A...389...58M}
{Mantovani}, F.; {Junor}, W.; {Ricci}, R.; {Saikia}, D.J.; {Salter}, C.;
  {Bondi}, M.
\newblock {Milli-arcsecond scale rotation measure in the CSS quasars 0548+165
  and 1524$-$136}.
\newblock {\em \aap} {\bf 2002}, {\em 389},~58--67.
\newblock {\url{https://doi.org/10.1051/0004-6361:20020482}}.

\bibitem[{Gaia Collaboration} et~al.(2023){Gaia Collaboration}, {Vallenari},
  {Brown}, {Prusti}, {de Bruijne}, {Arenou}, {Babusiaux}, {Biermann},
  {Creevey}, {Ducourant}, {Evans}, {Eyer}, {Guerra}, {Hutton}, {Jordi},
  {Klioner}, {Lammers}, {Lindegren}, {Luri}, {Mignard}, {Panem}, {Pourbaix},
  {Randich}, {Sartoretti}, {Soubiran}, {Tanga}, {Walton}, {Bailer-Jones},
  {Bastian}, {Drimmel}, {Jansen}, {Katz}, {Lattanzi}, {van Leeuwen}, {Bakker},
  {Cacciari}, {Casta{\~n}eda}, {De Angeli}, {Fabricius}, {Fouesneau},
  {Fr{\'e}mat}, {Galluccio}, {Guerrier}, {Heiter}, {Masana}, {Messineo},
  {Mowlavi}, {Nicolas}, {Nienartowicz}, {Pailler}, {Panuzzo}, {Riclet}, {Roux},
  {Seabroke}, {Sordo}, {Th{\'e}venin}, {Gracia-Abril}, {Portell}, {Teyssier},
  {Altmann}, {Andrae}, {Audard}, {Bellas-Velidis}, {Benson}, {Berthier},
  {Blomme}, {Burgess}, {Busonero}, {Busso}, {C{\'a}novas}, {Carry}, {Cellino},
  {Cheek}, {Clementini}, {Damerdji}, {Davidson}, {de Teodoro}, {Nu{\~n}ez
  Campos}, {Delchambre}, {Dell'Oro}, {Esquej}, {Fern{\'a}ndez-Hern{\'a}ndez},
  {Fraile}, {Garabato}, {Garc{\'\i}a-Lario}, {Gosset}, {Haigron}, {Halbwachs},
  {Hambly}, {Harrison}, {Hern{\'a}ndez}, {Hestroffer}, {Hodgkin}, {Holl},
  {Jan{\ss}en}, {Jevardat de Fombelle}, {Jordan}, {Krone-Martins}, {Lanzafame},
  {L{\"o}ffler}, {Marchal}, {Marrese}, {Moitinho}, {Muinonen}, {Osborne},
  {Pancino}, {Pauwels}, {Recio-Blanco}, {Reyl{\'e}}, {Riello}, {Rimoldini},
  {Roegiers}, {Rybizki}, {Sarro}, {Siopis}, {Smith}, {Sozzetti}, {Utrilla},
  {van Leeuwen}, {Abbas}, {{\'A}brah{\'a}m}, {Abreu Aramburu}, {Aerts},
  {Aguado}, {Ajaj}, {Aldea-Montero}, {Altavilla}, {{\'A}lvarez}, {Alves},
  {Anders}, {Anderson}, {Anglada Varela}, {Antoja}, {Baines}, {Baker},
  {Balaguer-N{\'u}{\~n}ez}, {Balbinot}, {Balog}, {Barache}, {Barbato},
  {Barros}, {Barstow}, {Bartolom{\'e}}, {Bassilana}, {Bauchet}, {Becciani},
  {Bellazzini}, {Berihuete}, {Bernet}, {Bertone}, {Bianchi}, {Binnenfeld},
  {Blanco-Cuaresma}, {Blazere}, {Boch}, {Bombrun}, {Bossini}, {Bouquillon},
  {Bragaglia}, {Bramante}, {Breedt}, {Bressan}, {Brouillet}, {Brugaletta},
  {Bucciarelli}, {Burlacu}, {Butkevich}, {Buzzi}, {Caffau}, {Cancelliere},
  {Cantat-Gaudin}, {Carballo}, {Carlucci}, {Carnerero}, {Carrasco},
  {Casamiquela}, {Castellani}, {Castro-Ginard}, {Chaoul}, {Charlot}, {Chemin},
  {Chiaramida}, {Chiavassa}, {Chornay}, {Comoretto}, {Contursi}, {Cooper},
  {Cornez}, {Cowell}, {Crifo}, {Cropper}, {Crosta}, {Crowley}, {Dafonte},
  {Dapergolas}, {David}, {David}, {de Laverny}, {De Luise}, and {De
  March}]{2023AA...674A...1G}
{Gaia Collaboration}.; {Vallenari}, A.; {Brown}, A.G.A.; {Prusti}, T.; {de
  Bruijne}, J.H.J.; {Arenou}, F.; {Babusiaux}, C.; {Biermann}, M.; {Creevey},
  O.L.; {Ducourant}, C.;  et~al.
\newblock {Gaia Data Release 3. Summary of the content and survey properties}.
\newblock {\em \aap} {\bf 2023}, {\em 674},~A1.
\newblock {\url{https://doi.org/10.1051/0004-6361/202243940}}.

\bibitem[{Fey} et~al.(2015){Fey}, {Gordon}, {Jacobs}, {Ma}, {Gaume}, {Arias},
  {Bianco}, {Boboltz}, {B{\"o}ckmann}, {Bolotin}, {Charlot}, {Collioud},
  {Engelhardt}, {Gipson}, {Gontier}, {Heinkelmann}, {Kurdubov}, {Lambert},
  {Lytvyn}, {MacMillan}, {Malkin}, {Nothnagel}, {Ojha}, {Skurikhina},
  {Sokolova}, {Souchay}, {Sovers}, {Tesmer}, {Titov}, {Wang}, and
  {Zharov}]{2015AJ....150...58F}
{Fey}, A.L.; {Gordon}, D.; {Jacobs}, C.S.; {Ma}, C.; {Gaume}, R.A.; {Arias},
  E.F.; {Bianco}, G.; {Boboltz}, D.A.; {B{\"o}ckmann}, S.; {Bolotin}, S.;
  et~al.
\newblock {The Second Realization of the International Celestial Reference
  Frame by Very Long Baseline Interferometry}.
\newblock {\em \aj} {\bf 2015}, {\em 150},~58.
\newblock {\url{https://doi.org/10.1088/0004-6256/150/2/58}}.

\bibitem[{Charlot} et~al.(2020){Charlot}, {Jacobs}, {Gordon}, {Lambert}, {de
  Witt}, {B{\"o}hm}, {Fey}, {Heinkelmann}, {Skurikhina}, {Titov}, {Arias},
  {Bolotin}, {Bourda}, {Ma}, {Malkin}, {Nothnagel}, {Mayer}, {MacMillan},
  {Nilsson}, and {Gaume}]{2020AA...644A.159C}
{Charlot}, P.; {Jacobs}, C.S.; {Gordon}, D.; {Lambert}, S.; {de Witt}, A.;
  {B{\"o}hm}, J.; {Fey}, A.L.; {Heinkelmann}, R.; {Skurikhina}, E.; {Titov},
  O.;  et~al.
\newblock {The third realization of the International Celestial Reference Frame
  by very long baseline interferometry}.
\newblock {\em \aap} {\bf 2020}, {\em 644},~A159.
\newblock {\url{https://doi.org/10.1051/0004-6361/202038368}}.

\bibitem[{Nothnagel} et~al.(2017){Nothnagel}, {Artz}, {Behrend}, and
  {Malkin}]{2017JGeod..91..711N}
{Nothnagel}, A.; {Artz}, T.; {Behrend}, D.; {Malkin}, Z.
\newblock {International VLBI Service for Geodesy and Astrometry: Delivering
  high-quality products and embarking on observations of the next generation}.
\newblock {\em J. Geod.} {\bf 2017}, {\em 91},~711--721.
\newblock {\url{https://doi.org/10.1007/s00190-016-0950-5}}.

\bibitem[{Mart{\'\i}-Vidal} et~al.(2016){Mart{\'\i}-Vidal}, {Roy}, {Conway},
  and {Zensus}]{2016A&A...587A.143M}
{Mart{\'\i}-Vidal}, I.; {Roy}, A.; {Conway}, J.; {Zensus}, A.J.
\newblock {Calibration of mixed-polarization interferometric observations.
  Tools for the reduction of interferometric data from elements with linear and
  circular polarization receivers}.
\newblock {\em \aap} {\bf 2016}, {\em 587},~A143.
\newblock {\url{https://doi.org/10.1051/0004-6361/201526063}}.

\bibitem[{Deller} et~al.(2011){Deller}, {Brisken}, {Phillips}, {Morgan},
  {Alef}, {Cappallo}, {Middelberg}, {Romney}, {Rottmann}, {Tingay}, and
  {Wayth}]{2011PASP..123..275D}
{Deller}, A.T.; {Brisken}, W.F.; {Phillips}, C.J.; {Morgan}, J.; {Alef}, W.;
  {Cappallo}, R.; {Middelberg}, E.; {Romney}, J.; {Rottmann}, H.; {Tingay},
  S.J.;  et~al.
\newblock {DiFX-2: A More Flexible, Efficient, Robust, and Powerful Software
  Correlator}.
\newblock {\em \pasp} {\bf 2011}, {\em 123},~275--287.
\newblock {\url{https://doi.org/10.1086/658907}}.

\bibitem[{Petit} and {Luzum}(2010)]{IERSTN36}
{Petit}, G.; {Luzum}, B., Eds.
\newblock {\em IERS Conventions}; { IERS Technical Note}; {Verlag
  des Bundesamts f\"{u}r Kartographie und Geod\"{a}sie}: Frankfurt am Main, Germany, 
  2010; Volume~36.

\bibitem[{Schaer} et~al.(1996){Schaer}, {Beutler}, {Rothacher}, and
  {Springer}]{Schaer1996}
{Schaer}, S.; {Beutler}, G.; {Rothacher}, M.; {Springer}, T.A.
\newblock Daily Global Ionosphere Maps Based on GPS Carrier Phase Data
  Routinely Produced by the CODE Analysis Center.
\newblock In \textit{Proceedings of the IGS 1996 Analysis Center Workshop}; {Neilan},
  R.; {Van Scoy}, P.; Zumberge, J., Eds.; Jet Propulsion Laboratory,
California Institute of Technology: Pasadena, CA, USA, 1996; pp.
  181--192.

\bibitem[{Aksim} et~al.(2019){Aksim}, {Melnikov}, {Pavlov}, and
  {Kurdubov}]{2019ApJ...885..159A}
{Aksim}, D.; {Melnikov}, A.; {Pavlov}, D.; {Kurdubov}, S.
\newblock {Exploring the Asymmetry of the Solar Corona Electron Density with
  Very Long Baseline Interferometry}.
\newblock {\em \apj} {\bf 2019}, {\em 885},~159.
\newblock {\url{https://doi.org/10.3847/1538-4357/ab499a}}.

\bibitem[{Titov} et~al.(2004){Titov}, {Tesmer}, and
  {Boehm}]{2004ivsg.conf..267T}
{Titov}, O.; {Tesmer}, V.; {Boehm}, J.
\newblock {OCCAM v.6.0 Software for VLBI Data Analysis}.
\newblock In \textit{Proceedings of the International VLBI Service for Geodesy and
  Astrometry 2004 General Meeting}; {Vandenberg}, N.R.; {Baver}, K.D., Eds.; NASA Goddard Space Flight Center: Greenbelt, MD, USA,
  2004; pp. 267--271.

\bibitem[{Petrov} and {Kovalev}(2025)]{2025ApJS..276...38P}
{Petrov}, L.Y.; {Kovalev}, Y.Y.
\newblock {The Radio Fundamental Catalog. I. Astrometry}.
\newblock {\em \apjs} {\bf 2025}, {\em 276},~38,
  \href{http://arxiv.org/abs/2410.11794}{{\normalfont
  [arXiv:astro-ph.IM/2410.11794]}}.
\newblock {\url{https://doi.org/10.3847/1538-4365/ad8c36}}.

\bibitem[{Shepherd} et~al.(1994){Shepherd}, {Pearson}, and
  {Taylor}]{1994BAAS...26..987S}
{Shepherd}, M.C.; {Pearson}, T.J.; {Taylor}, G.B.
\newblock {DIFMAP: An interactive program for synthesis imaging.}
\newblock {\em \baas} {\bf 1994}, {\em 26},~987--989.

\bibitem[{Beasley} et~al.(2002){Beasley}, {Gordon}, {Peck}, {Petrov},
  {MacMillan}, {Fomalont}, and {Ma}]{2002ApJS..141...13B}
{Beasley}, A.J.; {Gordon}, D.; {Peck}, A.B.; {Petrov}, L.; {MacMillan}, D.S.;
  {Fomalont}, E.B.; {Ma}, C.
\newblock {The VLBA Calibrator Survey-VCS1}.
\newblock {\em \apjs} {\bf 2002}, {\em 141},~13--21.
\newblock {\url{https://doi.org/10.1086/339806}}.

\bibitem[{Niell} et~al.(2018){Niell}, {Barrett}, {Burns}, {Cappallo}, {Corey},
  {Derome}, {Eckert}, {Elosegui}, {McWhirter}, {Poirier}, {Rajagopalan},
  {Rogers}, {Ruszczyk}, {SooHoo}, {Titus}, {Whitney}, {Behrend}, {Bolotin},
  {Gipson}, {Gordon}, {Himwich}, and {Petrachenko}]{2018RaSc...53.1269N}
{Niell}, A.; {Barrett}, J.; {Burns}, A.; {Cappallo}, R.; {Corey}, B.; {Derome},
  M.; {Eckert}, C.; {Elosegui}, P.; {McWhirter}, R.; {Poirier}, M.;  et~al.
\newblock {Demonstration of a Broadband Very Long Baseline Interferometer
  System: A New Instrument for High-Precision Space Geodesy}.
\newblock {\em Radio Sci.} {\bf 2018}, {\em 53},~1269--1291.
\newblock {\url{https://doi.org/10.1029/2018RS006617}}.

\bibitem[{Xu} et~al.(2021){Xu}, {Savolainen}, {Zubko}, {Poutanen}, {Lunz},
  {Schuh}, and {Wang}]{2021JGRB..12621238X}
{Xu}, M.H.; {Savolainen}, T.; {Zubko}, N.; {Poutanen}, M.; {Lunz}, S.; {Schuh},
  H.; {Wang}, G.L.
\newblock {Imaging VGOS Observations and Investigating Source Structure
  Effects}.
\newblock {\em J. Geophys. Res. (Solid Earth)} {\bf 2021}, {\em
  126},~e2020JB021238. 
\newblock
  {\url{ https://doi.org/10.1029/2020JB021238}}.

\bibitem[{Tingay} et~al.(2003){Tingay}, {Jauncey}, {King}, {Tzioumis},
  {Lovell}, and {Edwards}]{2003PASJ...55..351T}
{Tingay}, S.J.; {Jauncey}, D.L.; {King}, E.A.; {Tzioumis}, A.K.; {Lovell},
  J.E.J.; {Edwards}, P.G.
\newblock {ATCA Monitoring Observations of 202 Compact Radio Sources in Support
  of the VSOP AGN Survey}.
\newblock {\em \pasj} {\bf 2003}, {\em 55},~351--384.
\newblock {\url{https://doi.org/10.1093/pasj/55.2.351}}.

\bibitem[{de Vries} et~al.(2004){de Vries}, {Becker}, {White}, and
  {Helfand}]{2004AJ....127.2565D}
{de Vries}, W.H.; {Becker}, R.H.; {White}, R.L.; {Helfand}, D.J.
\newblock {Optical Properties of faint FIRST Variable Radio Sources}.
\newblock {\em \aj} {\bf 2004}, {\em 127},~2565--2578.
\newblock {\url{https://doi.org/10.1086/383550}}.

\bibitem[{Nyland} et~al.(2020){Nyland}, {Dong}, {Patil}, {Lacy}, {van Velzen},
  {Kimball}, {Sarbadhicary}, {Hallinan}, {Baldassare}, {Clarke}, {Goulding},
  {Greene}, {Hughes}, {Kassim}, {Kunert-Bajraszewska}, {Maccarone}, {Mooley},
  {Mukherjee}, {Peters}, {Petrov}, {Polisensky}, {Rujopakarn}, {Whittle}, and
  {Vaccari}]{2020ApJ...905...74N}
{Nyland}, K.; {Dong}, D.Z.; {Patil}, P.; {Lacy}, M.; {van Velzen}, S.;
  {Kimball}, A.E.; {Sarbadhicary}, S.K.; {Hallinan}, G.; {Baldassare}, V.;
  {Clarke}, T.E.;  et~al.
\newblock {Quasars That Have Transitioned from Radio-quiet to Radio-loud on
  Decadal Timescales Revealed by VLASS and FIRST}.
\newblock {\em \apj} {\bf 2020}, {\em 905},~74.
\newblock {\url{https://doi.org/10.3847/1538-4357/abc341}}.

\bibitem[{Komossa} et~al.(2026){Komossa}, {Grupe}, {Kraus}, {Edwards},
  {Kerrison}, {Rose}, {Soria}, {An}, {Hardcastle}, {Gab{\'a}nyi}, {Panda},
  {Xu}, {Wang}, {Frey}, and {Mez{\H{o}}si}]{2026ApJ..1003...46K}
{Komossa}, S.; {Grupe}, D.; {Kraus}, A.; {Edwards}, P.G.; {Kerrison}, E.F.;
  {Rose}, K.; {Soria}, R.; {An}, T.; {Hardcastle}, M.J.; {Gab{\'a}nyi},
  K.{\'E}.;  et~al.
\newblock {SDSS J110546.07+145202.4: The First Long-duration Radio
  Changing-look NLS1 Galaxy}.
\newblock {\em \apj} {\bf 2026}, {\em 1003},~46.
\newblock {\url{https://doi.org/10.3847/1538-4357/ae610f}}.

\bibitem[{Morganti}(2024)]{2024Galax..12...11M}
{Morganti}, R.
\newblock {What Have We Learned about the Life Cycle of Radio Galaxies from New
  Radio Surveys}.
\newblock {\em Galaxies} {\bf 2024}, {\em 12},~11.
\newblock {\url{https://doi.org/10.3390/galaxies12020011}}.

\bibitem[{Garofalo} and {Singh}(2019)]{2019ApJ...871..259G}
{Garofalo}, D.; {Singh}, C.B.
\newblock {FR0 Radio Galaxies and Their Place in the Radio Morphology
  Classification}.
\newblock {\em \apj} {\bf 2019}, {\em 871},~259.
\newblock {\url{https://doi.org/10.3847/1538-4357/aaf056}}.

\bibitem[{Garofalo} et~al.(2026){Garofalo}, {Liu}, and
  {Magerko}]{2026Galax..14....2G}
{Garofalo}, D.; {Liu}, Z.; {Magerko}, A.V.
\newblock {On the Origin and Nature of Double-Double Radio Galaxies}.
\newblock {\em Galaxies} {\bf 2026}, {\em 14},~2.
\newblock {\url{https://doi.org/10.3390/galaxies14010002}}.

\bibitem[{Garofalo} and {Singh}(2020)]{2020ApJ...905..156G}
{Garofalo}, D.; {Singh}, C.B.
\newblock {The Emergence of the X-Ray Luminosity/Cluster Richness Relation for
  Radio Galaxies}.
\newblock {\em \apj} {\bf 2020}, {\em 905},~156.
\newblock {\url{https://doi.org/10.3847/1538-4357/abc696}}.

\bibitem[{Best} et~al.(2003){Best}, {Lehnert}, {Miley}, and
  {R{\"o}ttgering}]{2003MNRAS.343....1B}
{Best}, P.N.; {Lehnert}, M.D.; {Miley}, G.K.; {R{\"o}ttgering}, H.J.A.
\newblock {Red galaxy overdensities and the varied cluster environments of
  powerful radio sources with z\raisebox{-0.5ex}\textasciitilde 1.6}.
\newblock {\em \mnras} {\bf 2003}, {\em 343},~1--21.
\newblock {\url{https://doi.org/10.1046/j.1365-8711.2003.06667.x}}.

\bibitem[Urunova et~al.(2025)Urunova, Kurdubov, and Melnikov]{Urunova2025}
Urunova, R.M.; Kurdubov, S.L.; Melnikov, A.E.
\newblock Change in the Position of the Radio Source 1328+254 in S and X-Bands.
\newblock {\em Mosc. Univ. Phys.} {\bf 2025}, {\em 80},~S143--S147.
\newblock {\url{https://doi.org/10.3103/S0027134925701590}}.



\end{thebibliography}
\end{document}